\PassOptionsToPackage{table,xcdraw}{xcolor}

\documentclass[review]{elsarticle}
\usepackage[utf8]{inputenc}

\usepackage{graphicx}
\usepackage{float}
\usepackage{amsmath}
\usepackage{mathtools}
\usepackage{listings}
\usepackage{array}
\usepackage{color}
\usepackage[table,xcdraw]{xcolor}
\usepackage[labelformat=simple]{subcaption}
\usepackage[inline]{enumitem}
\usepackage{multicol}
\usepackage{multirow}
\usepackage[ruled,vlined]{algorithm2e}
\usepackage[nolist]{acronym}
\usepackage{booktabs} 
\usepackage{balance}
\usepackage{hhline}
\usepackage{fancyhdr} 
\usepackage{hyperref}

\newcounter{todos}
\makeatletter
\renewcommand\p@subfigure{\thefigure\,}

\makeatother

\SetCommentSty{mycommfont}

\makeatletter
\newcommand{\mathleft}{\@fleqntrue\@mathmargin8pt}
\newcommand{\mathcenter}{\@fleqnfalse}
\makeatother

\makeatletter
\g@addto@macro\normalsize{%
  \setlength\abovedisplayskip{4pt}
  \setlength\belowdisplayskip{0pt}
  \setlength\abovedisplayshortskip{4pt}
  \setlength\belowdisplayshortskip{0pt}
}

\definecolor{pred}{rgb}{0.9,0,0}
\definecolor{pgrey}{rgb}{0.46,0.45,0.48}
\definecolor{javared}{rgb}{0.6,0,0} 
\definecolor{javagreen}{rgb}{0.25,0.5,0.35} 
\definecolor{javapurple}{rgb}{0.5,0,0.35} 
\definecolor{javablue}{rgb}{0.25,0.35,0.75} 
\definecolor{javabrown}{rgb}{0.41,0.24,0.24} 

\lstdefinestyle{walaIR}{language=SQL,
  showspaces=false,
  showtabs=false,
  breaklines=true,
  showstringspaces=false,
  breakatwhitespace=true,
  xleftmargin=4mm,
  basicstyle=\ttfamily\footnotesize,
  moredelim=[il][\textcolor{pgrey}]{\$\$},
  moredelim=[is][\textcolor{pgrey}]{\%\%}{\%\%},
  keywords={getfield,invokemethod,conditionalbranch,checkcast,goto},
  keywordstyle=\bfseries,
  emph=[1]{iterator, hasNext, next, getAccount, setCustomer},
  emphstyle=[1]{\color{javared}\bfseries},
  emph=[2]{transactions, manager, type, typeID},
  emphstyle=[2]{\color{javablue}\bfseries},
  frame=l,
  captionpos=b,
  belowskip=1.2em,
  aboveskip=1.2em,
  numbers=left,
  numberstyle=\tiny,
  mathescape,
  escapeinside={(*}{*)},
  backgroundcolor=\color{black!4!white}
}

\acrodef{CAPre}[CAPre]{CAPre}

\begin{document}

\title{CAPre: Code-Analysis based Prefetching for Persistent Object Stores\tnoteref{t1, t2}}
\tnotetext[t1]{DOI: \href{https://doi.org/10.1016/j.future.2019.10.023}{10.1016/j.future.2019.10.023}.\copyright Elsevier 2019. This manuscript version is made available under the \href{http://creativecommons.org/licenses/by-nc-nd/4.0/}{CC-BY-NC-ND 4.0 license}}

\author[1]{Rizkallah Touma}

\author[1]{Anna Queralt}

\author[1,2]{Toni Cortes} 

\address[1]{%
Barcelona Supercomputing Center, Jordi Girona 29, 08034 Barcelona}

\address[2]{%
Universitat Politècnica de Catalunya, Jordi Girona 31, 08034 Barcelona}



\begin{abstract}
Data prefetching aims to improve access times to data storage systems by predicting data records that are likely to be accessed by subsequent requests and retrieving them into a memory cache before they are needed. In the case of Persistent Object Stores, previous approaches to prefetching have been based on predictions made through analysis of the store's schema, which generates rigid predictions, or monitoring access patterns to the store while applications are executed, which introduces memory and/or computation overhead.

In this paper, we present \acs{CAPre}, a novel prefetching system for Persistent Object Stores based on static code analysis of object-oriented applications. \acs{CAPre} generates the predictions at compile-time and does not introduce any overhead to the application execution. Moreover, \acs{CAPre} is able to predict large amounts of objects that will be accessed in the near future, thus enabling the object store to perform parallel prefetching if the objects are distributed, in a much more aggressive way than in schema-based prediction algorithms. We integrate \acs{CAPre} into a distributed Persistent Object Store and run a series of experiments that show that it can reduce the execution time of applications from 9\% to over 50\%, depending on the nature of the application and its persistent data model.
\end{abstract}

%
%




\lstdefinestyle{code}{
  float,
  floatplacement=tbp
}

\maketitle

\section{Introduction}
\label{sec:intro}
Persistent Object Stores (POSs) are data storage systems that record and retrieve persistent data in the form of complete objects \cite{Brown1992}. They are especially used with Object-Oriented programming languages to avoid the impedance mismatch that occurs when developing OO applications on top of other types of databases, such as Relational Database Management Systems (RDBMSs). POSs make it easier to access persistent data without worrying about database access and query details, which can amount to 30\% of the total code of an application \cite{Atkinson1983,Chen2014ORM}.

Examples of POSs include object-oriented databases (e.g. Cach{\'e} \cite{cacheOODB} and Actian NoSQL \cite{actianNoSQL}) and Object-Relational Mapping (ORM) systems (e.g. Hibernate \cite{hibernateORM}, Apache OpenJPA \cite{apacheOpenJPA} and DataNucleus \cite{dataNucleus}). The rise of NoSQL databases has also led to the development of mapping systems for non-relational databases, such as Neo4J's Object-Graph Mapping (OGM) \cite{neo4jOgm}. Moreover, several POSs that support data distribution have been developed to accommodate the needs of parallel and distributed programming (e.g. Mneme \cite{Moss1990}, Nexus \cite{Tripathi1992}, Thor \cite{Barbara1999} and dataClay \cite{dataClayWeb,marti2017}).

Like in any other storage system, accessing persistent media is very slow and thus prefetching is needed to improve access times to stored data. Previous approaches to prefetching in POSs can be split into three broad categories:
\begin{enumerate*}
    \item schema-based,
    \item data-based, and
    \item code-based.
\end{enumerate*}
An example of a schema-based approach is the \emph{Referenced-Objects Predictor (ROP)}, which uses the following heuristic: each time an object is accessed, all the objects referenced from it are likely to be accessed as well \cite{hibernatePrefetching}. This type of approach gives rigid predictions that do not take into account how persistent objects are accessed by different applications. Nevertheless, ROP is widely used in commercial POSs because it achieves a reasonable accuracy and does not involve a costly prediction process (see Section ~\ref{sec:relatedWork}).

On the other hand, data-based approaches predict which objects to prefetch by detecting data access patterns while monitoring application execution. This type of approaches causes overhead that can amount to roughly 10\% of the application execution time \cite{garbatov2011data}. Furthermore, they may require large amounts of memory to store the detected patterns. Finally, few approaches have based the predictions on analyzing the source code of the OO applications that access the POS, and these have been largely theoretical without any in-depth analysis of the prediction accuracy or the performance improvement that they can achieve. For more details, Section \ref{sec:relatedWork} includes a study of the related work in the field of prefetching in POSs.

In this paper, we present an approach to predict access to persistent objects through static code analysis of object-oriented applications. The approach includes a complex inter-procedural analysis and takes non-deterministic program behavior into consideration. Then, we present \acs{CAPre}: a prefetching system that uses this prediction approach to prefetch objects from a POS. \acs{CAPre} performs the prediction at compile-time without adding any overhead to application execution time. It then uses source code generation and injection to modify the application's original code to activate automatic prefetching of the predicted objects when the application is executed. \acs{CAPre} also includes a further optimization by automatically prefetching data in parallel whenever possible, in order to maximize the benefits obtained from prefetching when using distributed POSs.

We integrate \acs{CAPre} into \emph{dataClay} \cite{marti2017}, a distributed POS, and run a series of experiments to measure the improvement in application performance that it can achieve. The experimental results indicate that using \acs{CAPre} to prefetch objects from a POS can reduce execution times of applications, with the most significant gains observed in applications with complex data models and/or many collections of persistent objects.

\subsubsection*{Contributions.}
The main contributions of the present paper can be summarized as follows:
\begin{itemize}
    \item We propose the theoretical basis of an approach to predict access to persistent objects based on static code analysis.
    \item We design and implement \acs{CAPre}, a prefetching system for Persistent Object Stores, using this prediction approach.
    \item We demonstrate how \acs{CAPre} improves the performance of applications by integrating it into an independent POS and running experiments on a set of well-known object-oriented and Big Data benchmarks.
\end{itemize}

The work reported here extends our previous work~\cite{Touma2017} in several directions. First, after presenting the theoretical grounds, we present the design and implementation of a complete prefetching system, based on static code analysis, and integrate it into an independent POS. Second, we evaluate the accuracy and performance gains obtained by our system by executing a set of benchmarks instead of simulating the expected accuracy results. These executions present the real effect of the technique on benchmarks and applications that were impossible to obtain by only using simulation.

\subsubsection*{Paper Organization}

Section \ref{sec:relatedWork} discusses the main differences of our proposal with current state of the art. Section \ref{sec:motivatingExample} presents an example that motivates our work and that will be used throughout the paper to guide the different steps. Section \ref{sec:formalization} summarizes the formalization of the used static code analysis approach. Section \ref{sec:systemOverview} presents our proposed prefetching system, \acs{CAPre}, and how it was implemented. Section \ref{sec:dataClay} discusses the integration details of \acs{CAPre} into a distributed POS. Section \ref{sec:evaluation} exposes the experimental evaluation of the system. Finally, Section \ref{sec:conclusions} concludes the paper and outlines some future work.

\section{Related Work}
\label{sec:relatedWork}

The structure in which Persistent Object Stores (POSs) expose data, in the form of objects and relations between these objects, is rich in semantics ideal for predicting access to persistent data and has invited a significant amount of research \cite{Knafla1997prefetching}. The most popular previous approach is the schema-based \emph{Referenced-Objects Predictor (ROP)}, defined in Section \ref{sec:intro}. Hibernate \cite{hibernatePrefetching}, Data Nucleus \cite{dataNucleusPrefetching}, Neo4JOGM \cite{neo4jOgm} and Spring Data JPA \cite{springPrefetching} all support this technique through specific configuration settings with varying degrees of flexibility (e.g. apply the prefetching on system level or only to specific types). For instance, Hibernate offers developers OR-Mappers \cite{Jboss}, which include predefined instructions that can be used to decide which related objects to prefetch for each object type, while with Django \cite{djangoPrefetching} developers need to supply explicit prefetching hints with each access to the POS. This type of implementation of ROP requires manual inspection of the entire application code by the developer and is an error-prone process, given that correct prefetching hints are difficult to determine and incorrect ones are hard to detect \cite{Ibrahim2006Automatic}. 

Schema-based techniques, as opposed to our proposal, only take into account the structure of the classes, but not how they are actually used by application methods, and thus can imply accessing a significant amount of unused data. Furthermore, given their heuristic nature, ROP approaches do not prefetch collections because the probability of bringing many unnecessary objects is very high. In our approach, as we will know exactly what collections will be accessed, we will show that we can use this information to prefetch them in a safe way increasing the effectiveness of the prefetching without incurring in unnecessary overhead.

Other prefetching mechanisms are data-based techniques that rely on the history of accesses to objects stored in the POS. Some examples of these approaches include object-page relation modeling \cite{Ahn2000SEOF,Knafla1997prefetching}, stochastic methods \cite{garbatov2011data}, Markov-Chains \cite{garbatov2011data,knafla1999analysing}, traversal profiling \cite{he2007path,Ibrahim2006Automatic}, the Lempel-Ziv data compression algorithm \cite{Curewitz1993} and context-aware prefetching \cite{bernstein1999context}. Moreover, predicting access to persistent objects at the type-level was first introduced by Han \emph{et al.} based on the argument that patterns do not necessarily exist between individual objects but rather between object types \cite{Han2005Formal}. The same authors later present an optimization of this approach by materializing the objects for each detected access pattern \cite{Han2006Type}. However, all of these approaches gather the information needed to make the predictions by monitoring access to the POS during application execution and thus introduce overhead in both memory and execution time.

Using code-based analysis to prefetch persistent objects was first suggested by Blair \emph{et al.}, who analyze the source code of OO applications at compile-time in order to model object relations and detect when the invocation of a method causes access to a different page \cite{blair2003classification}. This information is then used at runtime in order to prefetch the page once the execution of the corresponding method starts. The main difference with our approach is that they  are based on page granularity, thus bringing and keeping many objects that may not be necessary just because they reside in the same page. 

Finally, there is a completely different approach based on the queries executed over the data: "query rewriting". This mechanism is another type of optimization that can be used to prefetch objects. The idea behind this mechanism is to execute queries that are made more general to prefetch objects that might be relevant for future requests. Nevertheless, this again is based on heuristics and many unnecessary objects may be brought to the cache adding overhead and filling the cache with useless data. For more information, \cite{knafla1999prefetching} includes an extensive, albeit outdated, survey of different prefetching techniques while both \cite{Gerlhof1994multi} and \cite{blair2003classification} present taxonomies categorizing prefetching techniques in object-oriented databases.

In summary, our approach performs the prediction process at compile-time and produces type-level prefetching hints, combining the benefits of both types of approaches. The advantage of performing the prediction process at compile-time is the absence of overhead present in techniques which need information gathered at runtime. Similarly, type-level prediction is more powerful than its object-level counterpart and can capture patterns even when different objects of the same type are accessed. Moreover, information is not stored for each individual object which reduces the amount of used memory \cite{Han2003PrefetchGuide}. Finally, our approach also prefetches individual objects instead of entire pages of objects, which reduces the amount of memory occupied by other objects in the same page that might not necessarily be accessed.

\section{Motivating Example}
\label{sec:motivatingExample}

\begin{figure*}[t]
\centering
\includegraphics[width=\textwidth]{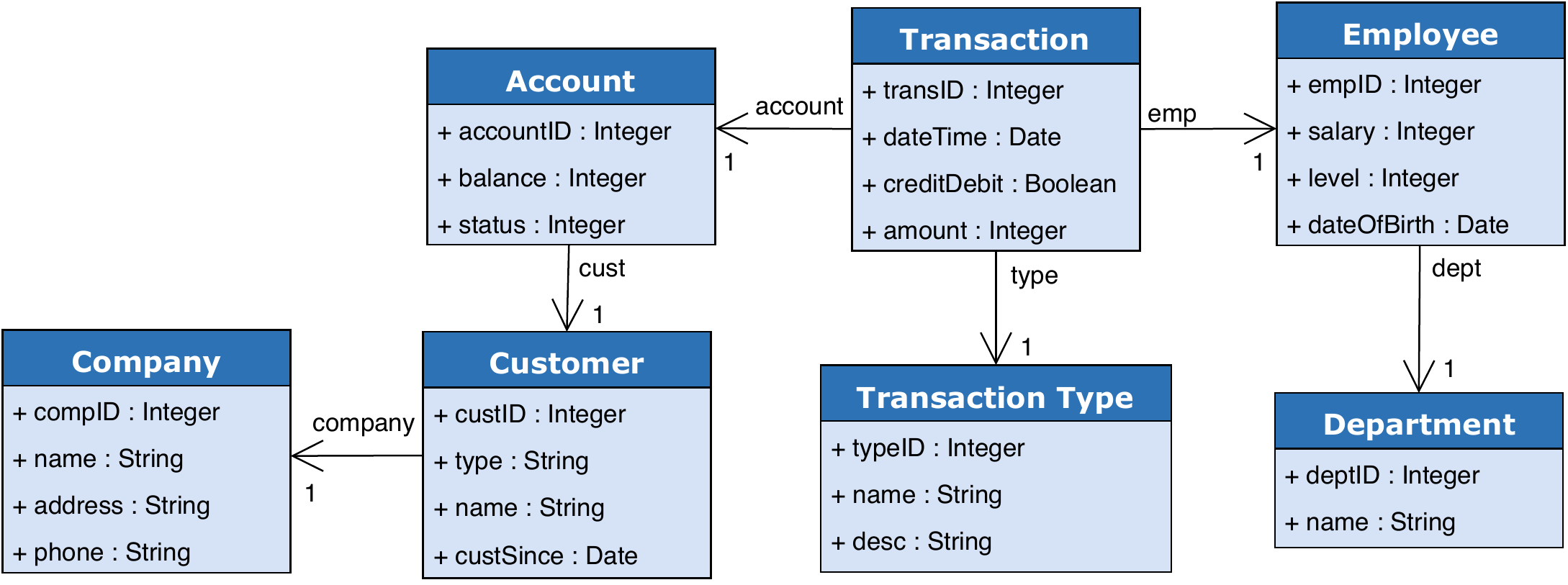}
\caption{Example of a Persistent Object Store (POS) schema. The schema represents a banking system with 7 entities, each of which corresponds to an object type in the POS.}
\label{fig:ex_motivating_example}
\end{figure*}

Figure \ref{fig:ex_motivating_example} shows the POS schema of a bank management system. In the figure, we can see various classes representing the entities of the system, such as \textit{Transaction}, \textit{Account} and \textit{Customer}. Let's say that we want to update the customers of the accounts responsible for all the transactions to be in the name of the manager of the bank. However, as a security measure, the system restricts updates on accounts to customers of the same company as the customer currently owning the account.

In order to achieve this task, we need to retrieve and iterate through all the \textit{Transaction} objects. We then navigate to the referenced \textit{Account} and \textit{Customer} until reaching the \textit{Company} of each customer. Finally, we need to compare the company of the customer currently owning the account with the company of the bank manager.

As we have mentioned, the most well used prediction technique that can be applied in this case is the  Referenced-Objects Predictor (ROP), defined in Section \ref{sec:intro}. Applying ROP to our example means that, for instance, each time a \textit{Transaction} object is accessed, the referenced \textit{Transaction Type}, \textit{Account} and \textit{Employee} objects are predicted to be accessed along with it.

However, in order to accomplish our task we also need to access the \textit{Customer} and \textit{Company} objects which will not be prefetched. On the other hand, the \textit{Transaction Type} and \textit{Employee} objects will be prefetched with \textit{Transaction} but in reality are not needed for the task at hand. To put this in numbers, if we have 100,000 \textit{Transactions} the ROP would wrongfully predict access to as many as 200,000 objects in the worst case while missing another 200,000 objects that will be accessed.

The prediction accuracy of ROP can be improved by increasing its "fetch depth", i.e. the number of levels of referenced objects to predict. For instance, instead of only predicting access to \textit{Transaction Type}, \textit{Account} and \textit{Employee}, which are directly referenced from \textit{Transaction}, having a fetch depth equal to 2 would also predict the objects referenced from them, which are \textit{Department} and \textit{Customer} in this example.

Increasing the fetch depth of ROP may help in predicting more relevant objects but it does not solve the problem of predicting access to objects that are not necessary. As a matter of fact, the more the fetch depth is increased the more likely it is to predict irrelevant objects as well. This is due to the fact that the ROP applies a heuristic based on the schema of the POS that does not take into account the application behavior.

Another more complex approach would be to monitor accesses to the POS and generate predictions based on the most commonly accessed objects \cite{Han2005Formal,Ibrahim2006Automatic,garbatov2011data}. For instance, monitoring accesses to the POS shown in Figure \ref{fig:ex_motivating_example} might tell us that in 80\% of the cases where a \textit{Transaction} object is accessed, its related \textit{Account} and \textit{Customer} objects are accessed as well.

This would work perfectly for our task, we will only need to load the referenced \textit{Company} object and all the other necessary objects will have been already prefetched. However, in the 20\% of cases where a transaction's \textit{Account} and \textit{Customer} are not needed, they will still be prefetched despite the fact that they will not be accessed. Moreover, retrieving the necessary information for this approach requires runtime monitoring of the application which adds overhead to the application execution time and memory consumption~\cite{garbatov2011data}.

The problem faced in both cases is that sometimes we prefetch objects that are not needed into memory and at the same time we don't prefetch objects that are actually accessed. This partially stems from the fact that the prediction heuristics are applied without taking into consideration the actual applications being used to access the data.
\section{Approach Formalization}
\label{sec:formalization}

\renewcommand{\ttdefault}{pcr}
\begin{lstlisting}[style=code, language=Java,
caption={Example OO application written in Java.},
captionpos=t,
label=ex:ooApplicationExample]
public class Transaction {
 private Account account;
 private Employee emp;
 private TransactionType type;

 public Account getAccount() {
  if (this.type.typeID == 1) {
   this.emp.doSmth();
  } else {
   this.emp.dept.doSmthElse();
  }
  return this.account;
 }
}

public class Account {
 private Customer cust;
 
 public void setCustomer(Customer newCust) {
   if (this.cust.company == newCust.company) {
     this.cust = newCust;
   }
 }
}

public class BankManagement {
 private ArrayList<Transaction> transactions;
 private Customer manager;

 public void setAllTransCustomers() {
   for (Transaction trans : this.transactions) {
     trans.getAccount().setCustomer(this.manager);
   }     
 }
}
\end{lstlisting}

This section summarizes the formalization of the approach we use to predict access to persistent objects. The formalization is based on the concept of \emph{type graphs} presented by Ibrahim and Cook~\cite{Ibrahim2006Automatic} that we have extended in order to capture the persistent objects accessed by a method in the form of a graph. After constructing these graphs, we generate a set of \emph{prefetching hints} that predict which objects should be prefetched from the POS for each method of the analyzed application.

\textbf{Example.} To help explain the approach, we use the sample object-oriented application shown in Listing \ref{ex:ooApplicationExample}, that uses the schema presented in Figure~\ref{fig:ex_motivating_example}, as a running example. 

\subsection{Initial definitions}
\label{sec:initialDefinitions}

For any such object-oriented application that uses a POS, we define $ T $ as the set of types of the application and $ PT \subseteq T $ as its subset of persistent types. Furthermore, $ \forall t \in T $ we define
\begin{itemize}
\item $ F_t:\text{ the set of persistent member fields of } t \text { such that } $ \\
$ \forall f \in F_t : type(f) \in PT $,

\item $ M_t:\text{ the set of member methods of } t $.
\end{itemize}

\subsection{Type Graphs}
\label{sec:typeGraphs}

\subsubsection{Application Type Graph} 
First, we need to represent in a graph all the relationships between classes in order to be able to decide which other classes are reachable starting from the fields of a given class. To keep this information, we represent the schema of the application through a directed type graph $ G_T = (T,A) $, where:
\begin{itemize}
\item $ T $ is the set of types defined by the application.
\item $ A $ is a function $ T \times F \rightarrow PT \times \left\{ single, collection \right\} $ representing a set of associations between types. Given types $ t $ and $ t' $ and field $ f $, if $ A(t, f) \rightarrow (t', c) $ then there is an association from $ t $ to $ t' $ represented by $ f \in F_t \text{ where } type(f) = t' $ with cardinality $ c $ indicating whether the association is \emph{single} or \emph{collection}.
\end{itemize}

\begin{figure}[!t]
\centering
\begin{subfigure}{.46\textwidth}
    \centering
	\includegraphics[width=0.95\textwidth]{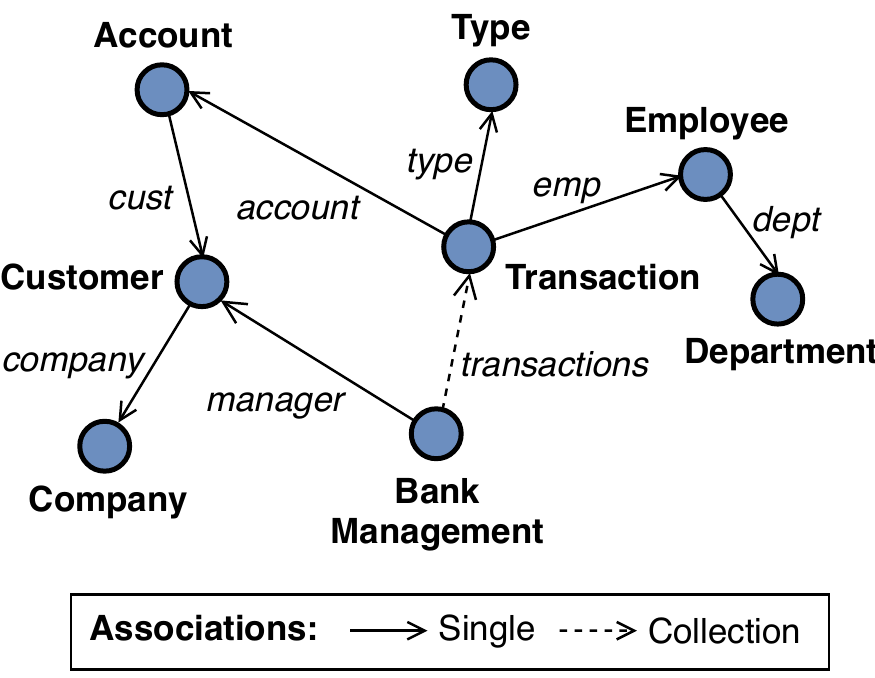}
	\caption{Type graph $ G_T $ of the whole application. }
	\label{fig:ex_type_graph}
\end{subfigure}\qquad
\begin{subfigure}{.46\textwidth}
    \centering
    \includegraphics[width=0.9\textwidth]{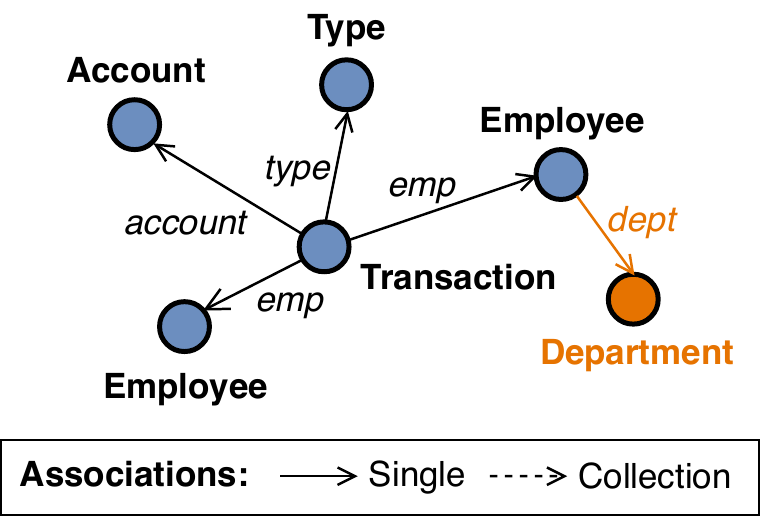}
    \caption{Type graph $ G_m $ of the method \textit{getAccount()}  (lines 6 to 13 from Listing \ref{ex:ooApplicationExample}). Branch-dependent navigations (Section \ref{sec:bdnom}) are highlighted in orange.}
    \label{fig:ex_method_type_graph}
\end{subfigure}%
\caption{Two type graphs from Listing \ref{ex:ooApplicationExample}. Solid lines represent single associations and dashed lines represent collection associations.}
\end{figure}

\textbf{Example.} Figure \ref{fig:ex_type_graph} shows the type graph of the application from Listing \ref{ex:ooApplicationExample}. Some of the associations of this type graph are:
\begin{itemize}
\item A(Bank Management, trans) $ \mapsto $ (Transaction, collection)
\item A(Transaction, account) $ \mapsto $ (Account, single)
\item A(Employee, dept) $ \mapsto $ (Department, single)
\end{itemize}


\subsubsection{Method Type Graph}
\label{sec:methodTypeGraph}
While $ G_T $ represents the general schema of the application, it does not capture how the associations between the different types are traversed by the application's methods. When a method $ m $ is executed, some of its instructions might trigger the navigation of a subset of the associations in $ G_T $.

An \emph{association navigation} $ t \rightharpoondown^{f} t' $ is triggered when an instruction accesses a field $ f $ in an object of type $ t $ (navigation source) to navigate to an object of type $ t' $ (navigation target) such that $ A(t, f) \rightarrow (t', c) $. A navigation of a collection association has multiple target objects corresponding to the collection's elements. The set of all association navigations in $ m $ form the method type graph $ G_m $, which is a sub-graph of $ G_T $ and captures the objects directly accessed by the method's instructions.

\textbf{Example.} Figure \ref{fig:ex_method_type_graph} shows the type graph $ G_m $ of the method \textit{getAccount()} with the implementation shown in Listing \ref{ex:ooApplicationExample} (lines 6 to 13). Notice that instructions that involve fields of primitive types, such as \textit{typeID} (integer), are not part of the graph because they do not trigger a navigation between objects.

\subsubsection{Augmented Method Type Graph}
The limitation of the method type graph ($ G_m $) is that it only includes association navigations that occur in the code of the method $ m $, but does not include associations navigated in other methods invoked by the original method $ m $. Thus, after constructing $ G_m $, we perform an inter-procedural analysis to capture the objects accessed inside other methods invoked by $ m $. The result of this analysis is the augmented method type graph $ AG_m $, which we construct by adding association navigations that are triggered inside an invoked method $ m' \in M_{t'} $ to $ G_m $ as follows:

\begin{itemize}
\item The type graph of the invoked method $ G_{m'} $ is added to $ G_m $ through the navigation $ t \rightharpoondown^{f} t' $ that caused the invocation.

\item Association navigations triggered by passing a persistent object as a parameter to $ m' $ are added directly to $ G_m $.
\end{itemize}

\begin{figure}[t]
\centering
\includegraphics[width=0.6\textwidth]{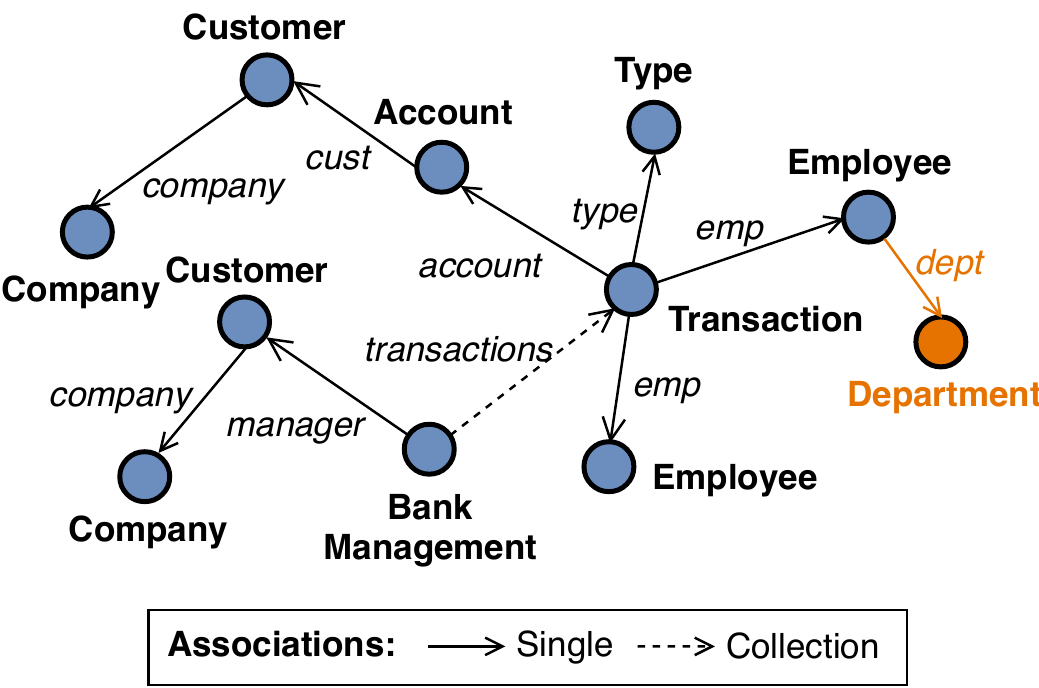}
\caption{Augmented method type graph $ AG_m $ of \emph{setAllTransCustomers()} from Listing \ref{ex:ooApplicationExample} (lines 30-34). Navigations highlighted in orange are branch-dependent (Section \ref{sec:bdnom}).}
\label{fig:ex_augmented_method_type_graph}
\end{figure}

\textbf{Example.} Figure \ref{fig:ex_augmented_method_type_graph} shows the augmented method type graph $ AG_m $ of method \emph{setAllTransCustomers()} from Listing \ref{ex:ooApplicationExample}. It includes the type graphs of the invoked methods \emph{getAccount()} and \emph{setCustomer(newCust)}. Note that the navigations $ BankManagement $ $ \rightharpoondown^{manager} Customer \rightharpoondown^{comp} Company $ are triggered by passing the persistent object \emph{BankManagement .manager} as a parameter to the method \emph{setCustomer(newCust)}.

\subsection{Prefetching Hints}
\label{sec:prefetchingHints}
After constructing the augmented type graph of a method, we can predict which objects will be accessed once the execution of the method starts. We achieve this by traversing $ AG_m $ and generating a set of prefetching hints $ PH_m $ that predict access to persistent objects:
\begin{itemize}[label={}]
    \item $ PH_m = \big\{ph\ |\ ph = f_1 . f_2 . \dotso . f_n 
\ \ where\ t_i \rightharpoondown^{f_i} t_{i+1} \in AG_M: 1 \leq i < n \big\}$
\end{itemize}
Each prefetching hint $ ph \in PH_m $ corresponds to a sequence of association navigations in $ AG_m $ and indicates that the target object(s) of the navigations is/are accessed.

\textbf{Example.} The augmented method type graph $ AG_m $ of Fig. \ref{fig:ex_augmented_method_type_graph} results in the following set of prefetching hints for method \emph{setAllTransCustomers()}. Note that hints starting with the collection \emph{transactions} predict that \emph{all} its elements are accessed:
\begin{itemize}[label={}]
    \item $ PH_m = \big\{transactions.type,\ transactions.emp, $ \\\hspace*{1.3cm}
    $ \ transactions.account.cust.company,\ manager.company \big\} $
\end{itemize}

\subsection{Runtime Application Behavior}
\label{sec:bdnom}

Given that we perform this analysis statically prior to the execution of the application, there are association navigations that we cannot decide if they are traversed or not, given that they depend on the runtime behavior of the application. Thus, in this section, we study how to react in such cases where a static analysis might lead to erroneous predictions of which objects should be prefetched. In particular, we considered two types of such behavior:
\begin{itemize}
    \item Navigations that depend on a method's branching behavior, which is determined by the method's conditional statements (e.g. \emph{if}, \emph{if-else}, \emph{switch-case}) and branching instructions (e.g. \emph{return}, \emph{break}). These navigations may or may not be triggered during execution, depending on which branch is taken, and hence might lead us to predict access to an object that does not occur. An example of this is $ Employee \rightharpoondown^{dept} Department $, highlighted in orange in Fig. \ref{fig:ex_augmented_method_type_graph}, which is only triggered inside the \emph{if} branch of a conditional statement.

    \item Navigations that are triggered inside a method's overridden versions. This behavior is caused by the dynamic binding feature of OO languages, which allows an object defined of type $ t $ to be initialized to a sub-type $ t' $. Thus, when invoking a method of type $ t $, the method being executed might actually be an overridden version defined in $ t' $, which in turn might result in erroneous predictions.
\end{itemize}

Once we have detected the problem, and before proposing a solution, we analyzed how often methods contain this kind of runtime-dependent behavior in order to understand the magnitude of the problem. We performed this analysis using the applications we will later use, in Section~\ref{sec:evaluation}, to evaluate our prefetching algorithm (OO7, WC, K-means, and PGA) combined with the applications of the SF110 corpus, which is a statistically representative sample of 100 Java applications from SourceForge, a popular open source repository, extended with the 10 most popular applications from the same repository~\cite{Gordon2014}.

\begin{figure}[!t]
\centering
\includegraphics[width=0.5\textwidth]{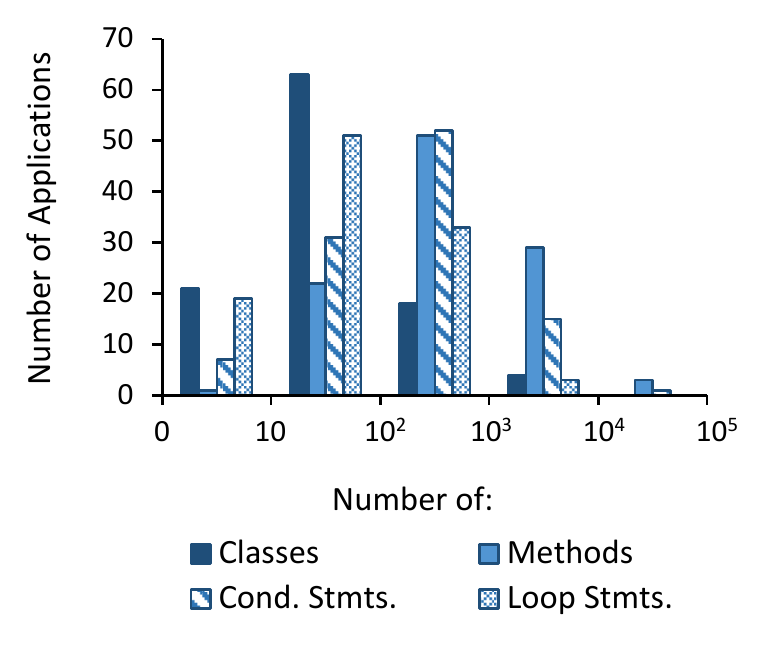}
\caption{For each power-of-10 interval on the x-axis, the y-axis represents the number of applications of the SF110 corpus that have the number of classes, methods, conditional statements and loop statements (as detected by our approach) in that interval. For instance, the first dark blue line starting from the left means that the number of applications that have between 0 and 10 methods is 20.}
\label{fig:eval_app_stats}
\end{figure}

Figure~\ref{fig:eval_app_stats} shows an aggregation of relevant characteristics of the applications used in our study: number of classes, methods, conditional statements and loop statements. Table \ref{tab:eval_apps_summary} also shows some summarized statistics of these characteristics and indicates that the test suite covers a wide range of applications, from very small applications to large applications containing over 20,000 methods.

\begin{table}[!t]
\caption{Summarized statistics of the corpus of applications used in our approach study.}
\label{tab:eval_apps_summary}
\centering
\small
\begin{tabular}{|lrrrr|r|}
\hline
~ & \multicolumn{1}{c}{\textbf{Max}} & \multicolumn{1}{c}{\textbf{Median}} & \multicolumn{1}{c}{\textbf{Avg}} & \multicolumn{1}{c|}{\textbf{Std. Dev.}} & \multicolumn{1}{c|}{\textbf{Total}} \\\hhline{======}
\# Classes & 2,292 & 38 & 139 & 381 & 14,760 \\
\# Methods & 26,261 & 335 & 1,379 & 3,517 & 146,182 \\
\# Cond. Stmts. & 17,935 & 162 & 656 & 1,893 & 69,495 \\
\# Loop Stmts. & 6,747 & 46 & 185 & 674 & 19,634 \\\hline
\end{tabular}
\end{table}

\begin{figure}[t]
\centering
    \begin{subfigure}[b]{0.49\textwidth}
        \centering
	    \includegraphics[width=\textwidth]{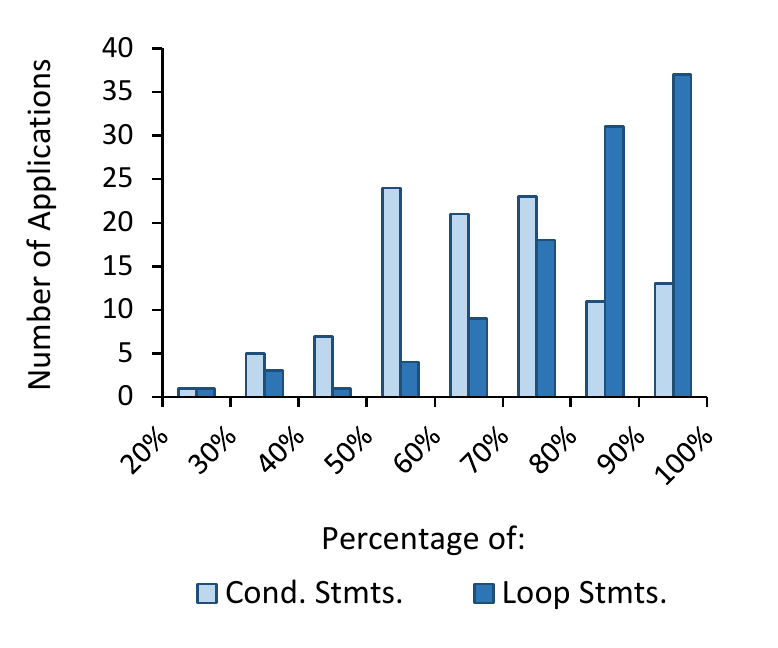}
        \caption{Conditional and Loop Statements}
        \label{fig:eval_branch_loop_stats}
    \end{subfigure}
    \begin{subfigure}[b]{0.49\textwidth}
        \centering
    	\includegraphics[width=\textwidth]{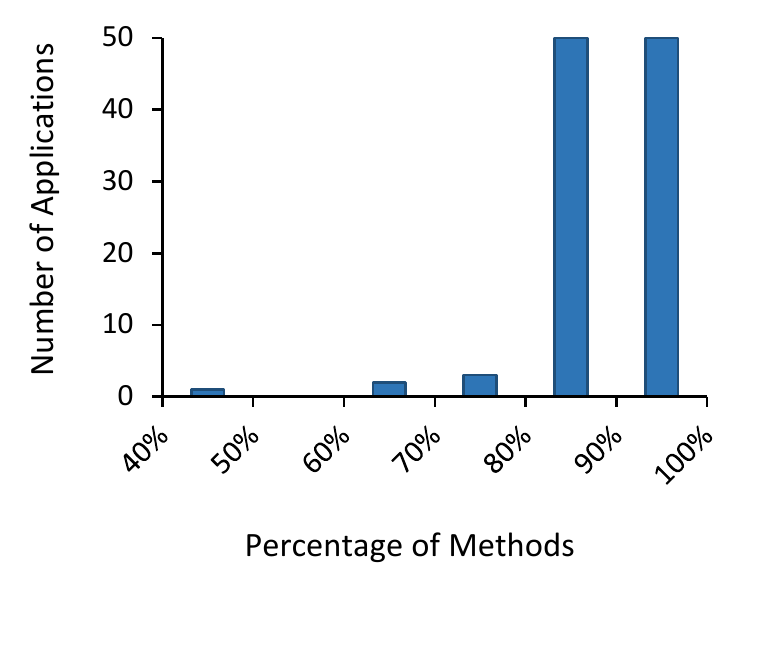}
        \caption{Methods}
        \label{fig:eval_method_stats}
    \end{subfigure}
\caption{For each 10\% interval on the x-axis, the y-axis represents the number of applications of the SF110 corpus that have the percentage of conditional statements, loop statements and methods that do not trigger any branch-dependent navigations in that interval.}
\label{fig:eval_stats}
\end{figure}

Let's now analyze the conditional and loop statements in the studied applications. Figure \ref{fig:eval_branch_loop_stats} shows the number of applications per percentage of conditional and loop statements that do not trigger any branch-dependent navigations. This means that the prefetching hints obtained when any branch is taken are the same (although the methods executed in each branch may be different, the accessed objects are the same). The category axis of Figure \ref{fig:eval_branch_loop_stats} starts at 20\% as none of the analyzed applications scored less in either case. It should be noted that one of the studied applications, \textit{greencow}, does not have any conditional statements while two, \textit{greencow} and \textit{dash-framework}, do not have any loop statements. Table \ref{tab:eval_stats_summary} shows that an average of 67.5\% of conditional statements and 82\% of loop statements do not trigger branch-dependent navigations, and hence do not pose a problem when generating access hints.

We aggregated these results to calculate the percentage of methods of each application that do not trigger any branch-dependent navigations, i.e. the methods for which our approach predicts the exact set of persistent objects that will be accessed. Figure \ref{fig:eval_method_stats} shows the results of this experiment, its category axis starts at 40\% as none of the studied applications scored a lower percentage. Figure \ref{fig:eval_method_stats} shows that only 6 of the studied applications scored below 80\%, which indicates that for 95.5\% of the studied applications, our approach can generate the exact set of access hints for over 80\% of methods. Table \ref{tab:eval_stats_summary} indicates that on average, 88.8\% of an application's methods do not trigger branch-dependent navigations, which is significantly higher than the average reported for conditional and loop statements, and also reports a low standard deviation of 7.9\%.

\begin{table}[!t]
\caption{Summarized statistics of the experimental results. The first three rows show the percentage of conditional statements, loop statements and methods that do not trigger any branch-dependent navigations. The last row shows the analysis time of the studied applications.}
\label{tab:eval_stats_summary}
\centering
\small
\begin{tabular}{|lrrrrr|}
\hline
~ & \multicolumn{1}{c}{\textbf{Min}} & \multicolumn{1}{c}{\textbf{Max}} & \multicolumn{1}{c}{\textbf{Median}} & \multicolumn{1}{c}{\textbf{Avg}} & \multicolumn{1}{c|}{\textbf{Std. Dev.}}  \\\hhline{======}
Cond. Stmts. (\%) & 26.8\% & 100\% & 67.1\% & 67.5\% & 17\%\\
Loop Stmts. (\%) & 24.8\% & 100\% & 85.7\% & 82\% & 15.7\% \\
Methods (\%) & 44\% & 100\% & 89.9\% & 88.8\% & 7.9\%\\
\hline
\end{tabular}
\end{table}

These results indicate that the prediction errors stemming from branch-dependent navigations are confined to a limited number of methods, while our static code analysis approach can accurately predict access to persistent objects in most cases. This is also in line with the intuition of the authors of \cite{Ibrahim2006Automatic} that accesses to persistent data are, in general, independent of an application's branching behavior.

Given these results, we conclude that the difference between the prefetching hints of the different branches of an application is quite small. Thus, in the implementation of \acs{CAPre} we will include hints corresponding to branch-dependent navigations (i.e. assuming both branches are taken) to increase the true positive rate (i.e. predicted objects that are accessed by the application), with minimal effect on false positives (i.e. predicted objects that are not accessed).

By contrast, our previous work has a detailed study indicating that including prefetching hints of overridden methods sharply increases the false positives rate in some cases \cite{Touma2017}. Based on the results of this study, in the implementation of \acs{CAPre}, we will not include the prefetching hints of overridden methods when generating $ PH_m $ of a particular method $ m $.
\section{System Overview}
\label{sec:systemOverview}

\begin{figure}[!t]
\centering
\includegraphics[width=0.6\textwidth]{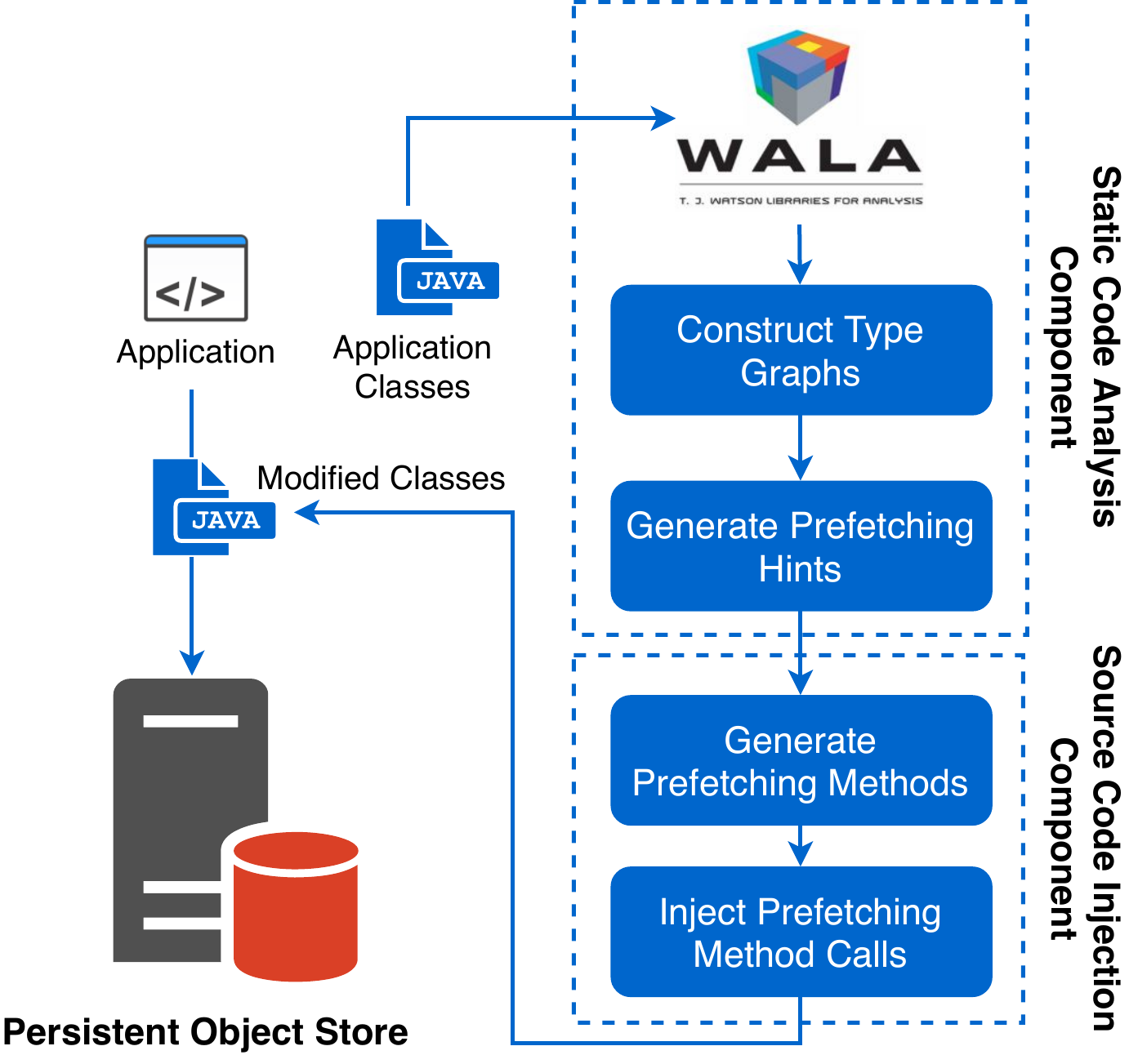}
\caption{Overview of the proposed prefetching system.}
\label{fig:system_overview}
\end{figure}

\acs{CAPre} is a prefetching system for Persistent Object Stores based on the static code analysis of object-oriented applications described in Section \ref{sec:formalization}. It consists of two main components, as depicted in Figure \ref{fig:system_overview}:
\begin{enumerate*}
    \item Static Code Analysis Component, and 
    \item Source Code Injection Component.
\end{enumerate*}
The \emph{Static Code Analysis Component} takes as input the source code of the application classes, written in Java, and executes the static analysis approach formalized in Section \ref{sec:formalization} in order to generate \emph{prefetching hints} that predict, for each method of the application, which persistent objects should be prefetched. We implemented this analysis for Java applications since it is the most common OO language, but the theoretical concepts of our approach can be applied to any other OO language.

Afterwards, the \emph{Source Code Injection Component} generates, for each method, a helper \emph{prefetching method} that prefetches the objects predicted by the generated prefetching hints. It also injects an invocation of this prefetching method to activate the prefetching automatically when the application is executed. The generated and injected code snippets uses multi-threading in order to perform the prefetching without interrupting the normal execution of the application, as well as to prefetch objects in parallel when using a distributed POS.

In the following subsections, we describe both components in detail.

\subsection{Static Code Analysis Component}
\label{sec:staticCodeAnalysis}
This component includes the implementation of the prediction approach summarized in Section \ref{sec:formalization}. We used IBM Wala \cite{ibmWala}, an open-source tool that parses and manipulates Java source code, to generate an Abstract Syntax Tree (AST) and an Intermediate Representation (IR) of each method of the analyzed application. We then constructed the augmented type graphs of the application's methods using these two structures, before finally generating the set of prefetching hints for each method.

\subsubsection{Wala AST and IR}

We used Wala's AST to identify conditional and loop statements. In particular, we identified two loop patterns used to iterate collections: using indexes or using iterators, each of which can be implemented with a \emph{for} or a \emph{while} loop. Similarly, we took \emph{if}, \emph{if-else} and \emph{switch-case} statements into consideration when identifying conditional statements. On the other hand, we used the IR, which contains a custom representation of the method's instructions, in order to detect association navigations that occur inside the method. Each IR instruction consists of five parts:
\begin{itemize}[topsep=2pt]
    \item $ II $: the instruction's index inside the IR,
    \item $ IType $: the instruction type (e.g. method invocation),
    \item $ IParams $: the instruction parameters (e.g. the invoked me\-thod, the accessed field),
    \item $ defVarId $: the ID of the variable defined by the instruction (can be null if the instruction doesn't define any variables),
    \item $ usedVarIDs[] $: zero or more previously-defined variables that are used by the instruction, indicated by their IDs.
\end{itemize}

\textbf{Example.} Listing \ref{ex:wala-ir} shows the IR instructions of the method \emph{setAllTransCustomers()} from Listing \ref{ex:ooApplicationExample}. The line numbers correspond to the instruction indexes (II). Note that $ II_2, II_3, II_4, II_5 $ and $ II_9 $ are implicit instructions generated due to the \emph{for} loop and are not explicitly invoked in the method's source code. Some examples of instructions from Listing \ref{ex:wala-ir} include:

\begin{itemize}
    \item $ II_1 $, $ IType = $ \emph{getfield}, $ IParams =~<$BankManagement, transactions, java/util/ArrayList$>,  defVarID = v_2 $, $ usedVarIDs $ $ = \{v_1\}$: this instruction accesses the field \emph{BankManagement.transactions} of type \emph{ArrayList} and assigns it the variable ID $ v_2 $. It also uses the variable ID $ v_1 $, which corresponds to the self-reference \emph{this}, to access the field.
    \item $ II_8 $, $ IType = $ \emph{invokemethod}, $ IParams =~<$ Account, setCustomer (Customer)V $>, defVarID = \phi $, $ usedVarIDs = \{v_6, v_7\}$: this instruction invokes the method \emph{Account .setCustomer(newCust)} and uses two variable IDs: $ v_6 $ corresponding to the object of type \emph{Account} on which the method is invoked, and $ v_7 $ corresponding to the field \emph{manager} used as a parameter of the invoked method.
\end{itemize}

\begin{lstlisting}[style=walaIR,
caption={Wala's Intermediate Representation (IR) of the method \emph{setAllTransCustomers()} from Listing \ref{ex:ooApplicationExample}.},
captionpos=t,
label=ex:wala-ir]
(*$ v_2 $*)= getfield <BankManagement, transactions, java/util/ArrayList>: (*$ v_1 $*)
(*$ v_3 $*)= invokemethod <java/util/ArrayList, iterator()java/util/(*Iterator*)>: (*$ v_2 $*)
(*$ v_4 $*)= invokemethod <java/util/(*Iterator*), hasNext()B>: (*$ v_3 $*)
conditionalbranch (eq, to iindex = -1): (*$ v_4,\ true $*)
(*$ v_5 $*)= invokemethod <java/util/(*Iterator*), next()java/lang/Object>: (*$ v_3 $*)
(*$ v_6 $*)= invokemethod <Transaction, getAccount()Account>: (*$ v_5 $*)
(*$ v_7 $*)= getfield <BankManagement, manager, Customer>: (*$ v_1 $*)
invokemethod <Account, setCustomer(Customer)V>: (*$ v_6, v_7 $*)
goto (from iindex = 10 to iindex = 3)
\end{lstlisting}

\subsubsection{Constructing Augmented Type Graphs}
Table \ref{tab:ssa_instrs_analysis} summarizes the IR instructions that we take into consideration when constructing the augmented method type graphs. We detect single association navigations with the instruction \emph{getfield} when the type of the used field is user-defined (i.e. the type corresponds to a class defined in the application). As for collection association navigations, we detect them when one of the two following instructions occur inside a loop statement:
\begin{itemize}
    \item \emph{arrayload}: which is used to access array elements,
    \item \emph{invokemethod} of the method \emph{next()} of the class \emph{java.util .Iterator}: which is used to access collection elements.
\end{itemize}

To detect branch-dependent navigations, we consider the bran\-ching instructions \emph{continue}, \emph{break} and \emph{return} when they occur inside a loop statement. When such an instruction is detected, the navigations resulting from all instructions inside the loop are marked as branch-dependent. Moreover, all navigations resulting from instructions inside a branch of a conditional statement are marked as branch-dependent.

We also use \emph{invokemethod} instructions to detect method invocations and augment the method's type graph with the type graph of the invoked method, as discussed in Section \ref{sec:typeGraphs}. When we do so, we bind the parameters of the method with the variables used in the invocation to detect association navigations triggered by passing a persistent object as a method parameter. Finally, we take into consideration \emph{return} instructions, if any, to detect the object that was returned by a method, which might be used to access further objects from the method invocation directy (e.g. \emph{getAccount().setCustomer(newCust)}).

\begin{table}[!t]
\caption{The IR instructions considered in our analysis.}
\label{tab:ssa_instrs_analysis}
\centering
\small
\begin{tabular}{lp{5.5cm}}
\toprule
\textbf{IR Instruction} & \textbf{Restrictions} \\\toprule
\multicolumn{2}{l}{\textbf{Single Association Navigations}} \\
getfield & User-defined field type \\\midrule
\multicolumn{2}{l}{\textbf{Collection Association Navigations}} \\
arrayload & Inside loop analysis scope \\
invokemethod & method \emph{java.util.Iterator.next()}, Inside loop statement \\\midrule
\multicolumn{2}{l}{\textbf{Branch-Dependent Navigations}} \\
break & Inside loop statement \\
continue & Inside loop statement \\
return & Inside loop statement \\\midrule
\multicolumn{2}{l}{\textbf{Method Invocations}} \\
invokemethod & Method of user-defined class \\\midrule
\multicolumn{2}{l}{\textbf{Method Return Object}} \\
return & \emph{N/A}\\
\bottomrule
\end{tabular}
\end{table}

These steps are detailed by the pseudo-code of Algorithm \ref{algo:construct_ag_m}, which takes as input the source code of a method $ m $ and returns as output the augmented method type graph $ AG_m $. The algorithm iterates through the instructions of the method and creates new nodes in $ AG_m $ through the method \emph{createNode()}, which takes as parameters the ID of the variable defined by the instruction, whether it corresponds to a navigation of a single or collection association and if it is branch-dependent. The method \emph{createEdge()} is used to add an edge to $ AG_m $ between the node of the current instruction and the nodes of previous instructions, based on the variable IDs used and defined by the instructions. Finally, in order to identify branch-dependent navigations, we implemented three helper methods used by the algorithm:
\begin{itemize}
    \item \emph{getASTNode(instr)}: which returns the AST node corresponding to an IR instruction, and
    \item \emph{hasConditionalParent(node)}, \emph{hasLoopParent(node)}: which indicate if an AST node has a parent node corresponding to a conditional or loop statement, respectively.
\end{itemize}

\begin{algorithm}
\caption{Construct Augmented Method Type Graph}
\label{algo:construct_ag_m}

\SetKwFunction{getASTNode}{getASTNode}
\SetKwFunction{hasConditionalParent}{hasConditionalParent}
\SetKwFunction{hasLoopParent}{hasLoopParent}
\SetKwFunction{type}{IType}
\SetKwFunction{params}{IParams}
\SetKwFunction{isParameterNode}{isParameterNode}
\SetKwFunction{usedVarIDs}{usedVarIDs}
\SetKwFunction{defVarID}{defVarID}
\SetKwFunction{getMethodGraph}{getMethodGraph}
\SetKwFunction{createNode}{createNode}
\SetKwFunction{createEdge}{createEdge}
\SetKwFunction{getNode}{getNode}
\SetKwFunction{bindParameter}{bindParameter}
\SetKwFunction{setIsReturnNode}{setIsReturnNode}

\SetKwInOut{Input}{Input}
\SetKwInOut{Output}{Output}

\Input{$ m \in M_t $: Source code of the method to analyze}
\Output{$ AG_m $: Augmented Method Type Graph of $ m $}
\BlankLine
$ AG_m \leftarrow (\phi, \phi) $\\
\BlankLine
\ForEach{instr $ \in I_m $} {
    instrASTNode $ \leftarrow $ \getASTNode(instr)
    \BlankLine
    \tcp{Identify branch-dependent navigations}
    \uIf{\hasConditionalParent(instrASTNode)) $ \vert \vert $
    (\hasLoopParent(instrASTNode)) \textup{\&\&}
    \type(instr) $ \in $ \{return, break, continue\})} {
        isBranchDependent $ \leftarrow $ true\\
    }
    \Else {
        isBranchDependent $ \leftarrow $ false\\
    }
    \BlankLine
    \tcp{Create single-association node in $ AG_m $}
    \If{\type(instr) = getfield \textup{\&\&} $ \params(instr).fieldType \in $ T } {
        $ AG_m \leftarrow AG_m \cup $ \createNode(\defVarID(instr), `single', isBranchDependent)\\
    }
    \BlankLine
    \tcp{Create collection-association node in $ AG_m $}
    \If{$ \big(( $\type(instr) = arrayload)
        $ \vert \vert $ (\type(instr) = invokemethod \textup{\&\&} \params(instr).invokedMethod = `java.util.Iterator.next()' $ )\big) $ \\
        \textup{\&\&} \hasLoopParent(instrASTNode)} {
        $ AG_m \leftarrow AG_m \cup $ \createNode(\defVarID(instr), `collection', isBranchDependent)\\
    }
    \BlankLine
    \tcp{Add nodes of invoked method to $ AG_m $}
    \If{\type(instr) = invokemethod \textup{\&\&} \params(instr).invokedMethod $ \in M_T $} {
        $ m' \leftarrow $ \params(instr).invokedMethod \\
        $ AG_{m'} \leftarrow $ \getMethodGraph($ m' $)\\
        \ForEach{node $ \in AG_{m'} $} {
            \uIf{\isParameterNode(node)}{
                $ AG_m \leftarrow AG_m \cup $ \bindParameter(node)\\
            }
            \Else {
                $ AG_m \leftarrow AG_m \cup $ node\\
            }
        }
    }
    \BlankLine
    \tcp{Flag return object of method}
    \If{\type(instr) = return} {
        usedNode $ \leftarrow $ getNode(\defVarID(instr))\\
        \setIsReturnNode(usedNode)\\
    }
    \BlankLine
    \tcp{Create edges between new and previous nodes}
    definedNode $ \leftarrow $ \getNode(\defVarID(instr))\\
    \ForEach{usedVarID $ \in $ \usedVarIDs(instr)} {
        usedNode $ \leftarrow $ \getNode(usedVarID)\\
        $ AG_m \leftarrow AG_m \cup $ \createEdge(usedNode, definedNode)\\
    }
}
\Return $ AG_m $
\end{algorithm}

\textbf{Example.} Applying Algorithm \ref{algo:construct_ag_m} on the instructions of \emph{setAllTransCustomers()} shown in Listing \ref{ex:wala-ir} results in the type graph $ AG_m $ depicted in Figure \ref{fig:ex_augmented_method_type_graph} as follows:

\begin{itemize}
    \item The instruction $ II_1 = $ \emph{getfield transactions} accesses a field of type \emph{collection}. Hence, no changes are made to $ AG_m $.
    \item $ II_5 $ is an invocation of \emph{java.util.Iterator.next()} inside a loop statement, which means it is accessing elements of the collection \emph{transactions}. Hence, a new node with the variable ID of \emph{transactions}, cardinality \emph{collection} and \emph{isBranchDepedent = false} is added to $ AG_m $.
    \item $ II_6 $ is an invocation of \emph{getAccount()}. Hence, the type graph of \emph{getAccount()} is added to $ AG_m $ and linked to the node corresponding to $ II_5 $, based on the used variable ID $ v_5 $.
    \item $ II_7 $ is a \emph{getfield} instruction that accesses the object \emph{manager}. Thus, it results in the creation of a new node with the variable ID of \emph{manager} and cardinality \emph{single}.
    \item $ II_8 $ is an invocation of \emph{setCustomer(newCust)} and results in adding its type graph to $ AG_m $, linking it to the node resulting from $ II_6 $, which represents the return object of \textit{getMethod()}. We also bind the method's parameter to the node resulting from $ II_7 $.
\end{itemize}

Note that $ II_2, II_3, II_4, \text{ and } II_9 $ do not access any persistent objects and hence do not cause any changes to $ AG_m $.

\subsubsection{Generating Prefetching Hints}

We generate the set of prefetching hints of a method $ PH_m $ by traversing the augmented method type graph constructed following Algorithm \ref{algo:construct_ag_m}. At this point, it is important to remember how we handle runtime application behavior (discussed in Section \ref{sec:bdnom}). In case of branch-dependent navigations, we will include the prefetching hints of all the branches, both taken and not taken, since it was shown in Section~\ref{sec:bdnom} to be the best option. On the other hand, we will not include any hints to prefetch the objects accessed by overridden methods, because it was shown to be a significant source of false positives in our previous work~\cite{Touma2017}.

We then perform one final modification to $ PH_m $ by removing hints already found in previous method calls. For instance, a method $ m $ that invokes another method $ m' $ will have the prefetching hints resulting from both $ m $ and $ m '$, which allows us to bring the prefetching forward ensuring that the predicted objects are prefetched before they are accessed.

However, this also means that $ m $ and $ m' $ might have prefetching hints predicting access to the same objects, which leads to launching several requests to prefetch the same objects when the application is executed, causing additional unnecessary overhead. We solve this problem by removing from $ PH_m $ those prefetching hints that are found in \emph{all} of the methods that invoke $ m $. This solution does not affect the prediction accuracy of the approach since the objects predicted by the removed hints are prefetched by other hints in a previously executed method.

\subsubsection{Computational Complexity of the Static Code Analysis}

Considering an application with a set of methods $ M $, Algorithm \ref{algo:construct_ag_m} has a complexity of $ O(|I_m|) $ when generating the augmented method type graph of any method $ m \in M $, where $ |I_m| $ is the number of Wala IR instructions of $ m $. Moreover, constructing the augmented method type graphs of all of the methods in $ M $ has a computational complexity of $ O(|M| * max(|I_m|)) $, where $ max(|I_m|) $ is the number of IR instructions of the largest method in the application.

This is due to the fact that each method of the application is only analyzed and its prefetching hints are only generated once, even if it is invoked multiple times by different methods of the application. Apart from this theoretical computational complexity, we provide detailed results of the time it takes to execute this static code analysis on various application in Section \ref{sec:Static-code-analysis-time}.

\subsection{Source Code Injection Component}
\label{sec:sourceCodeGeneration}
The goal of this component is to modify the original source code of the application in order to prefetch the objects predicted by the prefetching hints generated by the Static Code Analysis Component. To do so, we first generate a helper \emph{prefetching method} for each method of the application, which loads the objects predicted by the method's prefetching hints from the POS. Afterwards, we use AspectJ to inject an invocation of the generated prefetching method inside each method of the application. By doing so, the objects predicted by a method's $ AG_m $ are automatically prefetched when the application is executed.

\subsubsection{Generating Prefetching Methods}
Given that each POS has specific instructions that are used to retrieve stored objects, the exact instructions used in the prefetching methods to load the predicted objects depend on the used POS. For the purposes of this example, we assume that the POS has an instruction called \emph{load()} that loads and returns a typed object from the POS. The generated prefetching method takes as parameter the object on which the original method is executed, starting from which it then prefetches the predicted objects.

\textbf{Example.} The Source Code Injection Component generates the following prefetching method for the method \emph{setAllTransCustomers()} from Listing \ref{ex:ooApplicationExample}. Note that the prefetching method is defined in a new prefetching class corresponding to the class \emph{BankManagement}. Also note that the instruction \emph{load()} is substituted with the concrete instruction that loads an object depending on the used POS, as will be explained in Section \ref{sec:dataClay}.
\begin{lstlisting}[style= code, language=Java,
caption={Helper prefetching method of \emph{setAllTransCustomers()} from Listing \ref{ex:ooApplicationExample}.},
captionpos=t,
label=ex:prefetchingMethod]
public class BankManagement_prefetch {
  public void setAllTransCustomers_prefetch (BankManagement rootObject) {
    for (Transaction trans : rootObject.load(transactions)) {
      trans.load(type);
      trans.load(emp);
      trans.load(account).load(cust).load(company);
    });
    rootObject.load(manager).load(company);
  }
}
\end{lstlisting}

\subsubsection{Enabling parallel prefetching}
\label{sec:parallelPrefetching}

We further optimize \acs{CAPre} by performing parallel prefetching when an application accesses objects stored in a distributed POS. For instance, in the set of prefetching hints $ PH_m $ defined in Section \ref{sec:formalization}, the elements of the \emph{transactions} collection can be prefetched in parallel if they are stored in different nodes of a distributed POS. On the other hand, distributing single-association hints, such as \emph{manager.company}, is not possible since we need to load the object \emph{manager} before its associated \emph{company} is loaded.

We implemented this parallel prefetching by using the Parallel Streams of Java 8, which convert a collection into a stream and divide it into several substreams. The Java Virtual Machine (JVM) then uses a predefined pool of threads to execute a specific task for each substream, which avoids the costs of creating and destroying threads in each prefetching method. The number of threads in the pool is set by JVM to the number of processor cores of the current machine and the management of the threads is done automatically by the JVM.

\textbf{Example.} The parallel version of the prefetching method \emph{setAllTransCustomers\_prefetch()} is shown in Listing \ref{ex:parallelPrefetchingMethod}.
\begin{lstlisting}[style=code, language=Java,
caption={Parallelized prefetching method of \emph{setAllTransCustomers()} from Listing \ref{ex:ooApplicationExample}.},
captionpos=t,
label=ex:parallelPrefetchingMethod]
public class BankManagement_prefetch {
  public static void setAllTransCustomers_prefetch (BankManagement rootObject) {
    // Parallel prefetching of collection elements
    rootObject.load(transactions).parallelStream().forEach(trans -> {
      trans.load(type);
      trans.load(emp);
      trans.load(account).load(cust).load(company);
    });
    // Cannot be parallelized
    rootObject.load(manager).load(company);
  }
}
\end{lstlisting}

\subsubsection{Injecting Prefetching Method Invocations}
\label{sec:invokingPrefetchingMethods}
Instead of directly invoking the prefetching methods, we implemented a multi-threaded approach where the prefetching methods are executed by a background thread in parallel to the main thread of the application. By doing so, we allow the execution of the application to continue uninterrupted while prefetching objects in another thread whenever possible.

We achieved this by using a thread pool executor that creates a pool of one or more threads at the application level and then schedules tasks for execution in the created threads. This solution helps to save resources, since threads are not created and destroyed multiple times, and also contains the parallelism in predefined limits, such as the number of threads that are run in parallel. Hence, we inject the following instruction into the class that contains the main method from which the execution of the application starts:
\begin{lstlisting}[ language=Java,numbers=none]
public static final ThreadPoolExecutor prefetchingExecutor = (ThreadPoolExecutor) Executors.newFixedThreadPool(1);
\end{lstlisting}

This instruction creates a thread pool executor with a single thread to execute the generated prefetching methods. Afterwards, we inject at the beginning of each method a scheduling of its helper prefetching method using this constructed thread pool. The executor then checks the scheduled tasks and executes them consecutively in its thread. Note that when using the parallel prefetching methods, the single thread of the executor creates multiple sub-threads to perform the prefetching in parallel.

\textbf{Example.} Listing \ref{ex:modifiedMethod} shows the injected instructions into the method \emph{setAllTransCustomers()}, which schedule its helper prefetching method \emph{setAllTransCustomers\_prefetch()} for execution.
\begin{lstlisting}[style=code, language=Java,
caption={Injected scheduling of the prefetching method from Listing \ref{ex:parallelPrefetchingMethod} into \emph{setAllTransCustomers()}.},
captionpos=t,
label=ex:modifiedMethod]
public void setAllTransCustomers() {
  // Injected scheduling of prefetching method
  final BankManagement rootObject = this; 
  prefetchingExecutor.submit(new Runnable() {
    @Override
    public void run() {
      BankManagement_prefetch .setAllTransCustomers_prefetch(rootObject);
    }
  });
  ...
}
\end{lstlisting}
\section{Prefetching in dataClay}
\label{sec:dataClay}
\begin{figure*}[!t]
\centering
\includegraphics[width=\textwidth]{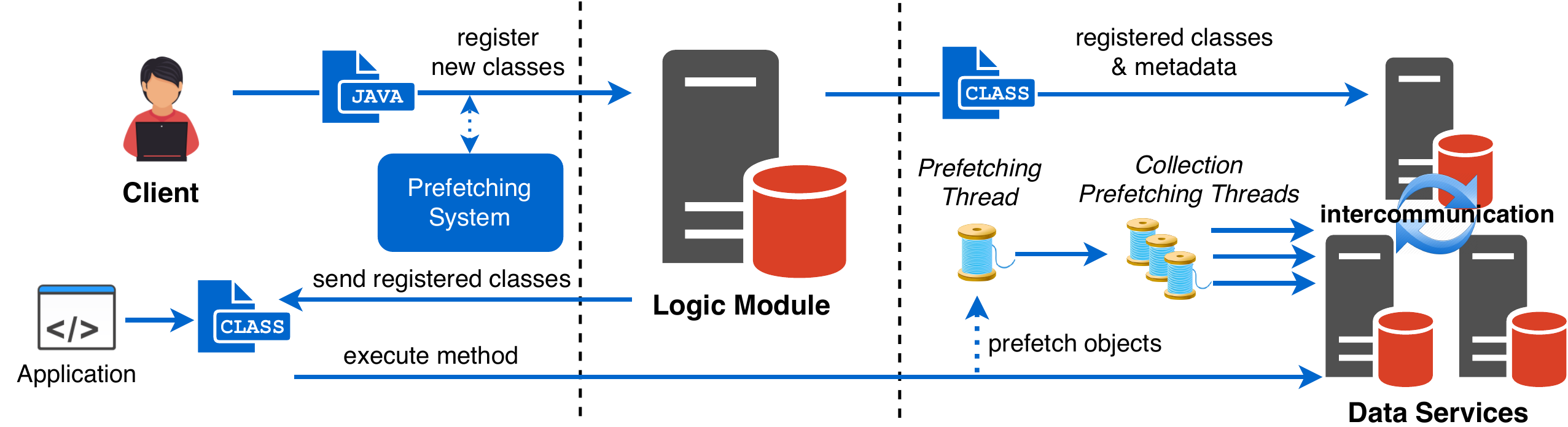}
\caption{System architecture of \emph{dataClay}. A deployment of a Logic Module and three Data Services on different nodes is depicted with the communications between the client and \emph{dataClay} and between Logic Module and Data Services \cite{marti2017}.}
\label{fig:dataClay_overview}
\end{figure*}

In order to evaluate the effect of \acs{CAPre} on application performance, we integrated it into \emph{dataClay}. \emph{dataClay} is an object store that distributes objects across the network \cite{marti2017,dataClayWeb} among the available storage nodes. In contrast with other database systems, data stored in \emph{dataClay} never moves outside the POS. Instead, data is manipulated in the form of objects, exposing only the operations that can be executed on the data, which are executed inside the data store, in a manner transparent to the applications using the store. Figure \ref{fig:dataClay_overview} shows the system architecture of \emph{dataClay}.

To use \emph{dataClay}, the client first needs to register the application schema, i.e. the set of persistent classes (fields and methods) that will be used by the application, to a centralized service called the \emph{Logic Module}. The Logic Module then adds system-specific functionality to the received classes and deploys the modified classes to the \emph{Data Services}, which are the nodes of \emph{dataClay} where the persistent objects are stored, and sends them back to the client.

We integrated \acs{CAPre} into \emph{dataClay} during this registration process. When the classes are sent to the Logic Module for registration, \acs{CAPre} intercepts the source code of the classes, performs the analysis and injects the prefetching classes and prefetching method invocations. These prefetching classes are then sent along with the modified application classes to the Logic Module for registration. Since \emph{dataClay} automatically loads an object when a reference to that object is made, the generated prefetching methods do not use any specific instructions to load the predicted objects but rather make explicit references to them (e.g. \emph{trans.type, trans.account.cust.company}).

Once the application schema is registered, the client can store any local objects with the type of a registered class in \emph{dataClay}, which automatically distributes the stored objects among the available Data Services. The client can then access the stored objects to execute any method defined in the registered schema. However, \emph{dataClay} does not send the objects to the client but rather executes the methods locally in the same Data Service where the object is stored.

Given the changes made by \acs{CAPre} during the schema registration, the helper prefetching method of the executed method is invoked once an execution request is received by a Data Service, and the predicted objects are prefetched into the local memory of the Data Service. When a prefetching method encounters an object in another Data Service, \textit{dataClay} communicates with that Data Service to load the object where it is stored.

\textbf{Example.} Executing the method \emph{setAllTransCustomers()} (Listing \ref{ex:modifiedMethod}) from a client application using \emph{dataClay} with three Data Services, $ DS_1 $, $ DS_2 $ and $ DS_3 $ (Figure \ref{fig:dataClay_overview}) on an object of type \emph{BankManagement} stored in $ DS_1 $, is done through the following steps:

\begin{itemize}
    \item First, the client application launches the execution request to \emph{dataClay}, which in turn automatically redirects it to $ DS_1 $, where the object \emph{BankManagement} is stored.
    \item When $ DS_1 $ receives the execution request of \emph{setAllTransCustomers()}, it schedules the prefetching method \emph{setAllTransCustomers\_prefetch()} for execution with the prefetching thread pool, as explained in Section \ref{sec:invokingPrefetchingMethods}.
    \item Once the prefetching method is executed, it creates several sub-threads and starts loading the elements of the collection \emph{transactions}, which was automatically distributed by \emph{dataClay}, in parallel from the different Data Services.
    \item When one of these threads, currently being executed on $ DS_1 $, tries to load an object stored in a different Data Service, say $ DS_2 $, \emph{dataClay} redirects the load request to $ DS_2 $ and loads the object where it is stored.
\end{itemize}
\section{Evaluation}
\label{sec:evaluation}

The purpose of this evaluation is to analyse how \acs{CAPre} reduces application execution time, which is the ultimate goal of our prefetching technique. For other indicators such as the true positive or the false negative rates, we refer the reader to~\cite{Touma2017}, where these metrics were analysed in detail. 

\subsection{Static-code analysis time}
\label{sec:Static-code-analysis-time}

Before we evaluate the performance gains obtained by applications when using \acs{CAPre}, it is important to prove that the proposed static code analysis and the generation of the prefetching hints can be run in a reasonable amount of time. In order to understand this, we have run the static code analysis using the applications of the SF110 corpus (introduced in Section~\ref{sec:bdnom}) as well as the applications we used to evaluate the performance gains of \acs{CAPre}, as detailed in Section~\ref{sec:performance}. 

\begin{figure}[!t]
\centering
\includegraphics[width=0.6\textwidth]{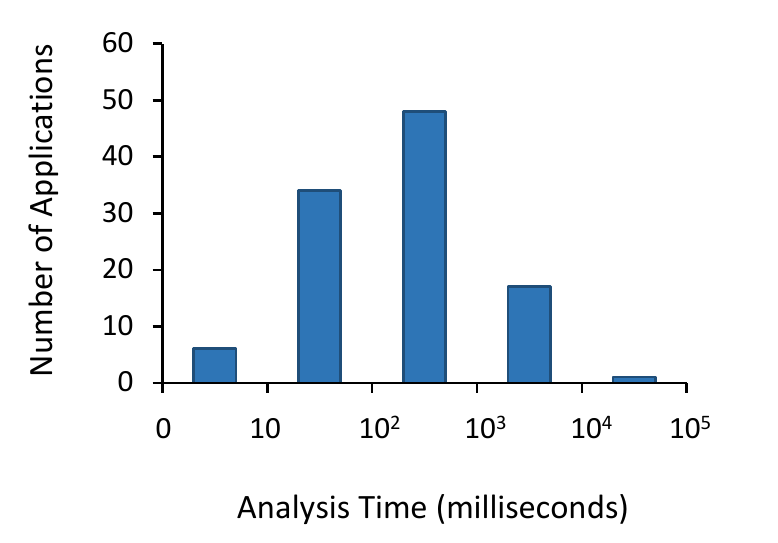}
\caption{For each power-of-10 interval on the x-axis, the y-axis represents the number of applications for which our static code analysis approach finishes within that interval (in milliseconds).}
\label{fig:eval_exec_time}
\end{figure}

Figure \ref{fig:eval_exec_time} plots the number of applications per range of analysis time in milliseconds and shows that 96 of the SF110 application were analyzed in less than 1 second. Moreover, it also shows that the longest time the static code analysis took was 16 seconds, and this occurred with \textit{weka}, the second largest application with over 20,000 methods.


As expected, the analysis time of our approach is correlated with the number of classes and methods of an application. However, with an average analysis time of 651 milliseconds and a maximum of roughly 16 seconds, we believe that the analysis finishes within a reasonable time for all of the analyzed applications. It is worth mentioning again here that this static analysis is done only once, prior to application execution and does not add any overhead to its execution time.

Going into more details of the four benchmarks that we will later use to asses the performance gains of \acs{CAPre}, Table \ref{tab:eval_compilation_analysis_time} shows the time needed to compile each of the benchmarks (by executing a \textit{javac} command) and the time needed to perform our code analysis. As we can see, the time needed to analyze the application code never exceeds the pure compilation time of the application, thus it will not imply a significant overhead when compiling the application (again, this analysis is only performed once before the applications are executed).

\begin{table}[!t]
\caption{Comparison between the compilation times and the times needed to perform the \acs{CAPre} static code analysis of each of the four benchmarks used in our evaluation (Section \ref{sec:performance}).}
\label{tab:eval_compilation_analysis_time}
\centering
\small
\begin{tabular}{|lrr|}
\hline
 & Compilation & \acs{CAPre} Analysis \\
\hline
\hline
OO7 & 1,030 ms & 827 ms\\
Wordcount & 923 ms & 633 ms\\
K-Means & 916 ms & 519 ms\\
PGA & 1,041 ms & 1,068 ms\\
\hline
\end{tabular}
\end{table}

\subsection{Evaluation of Application Performance}
\label{sec:performance}

We tested the effect that \acs{CAPre} has on application performance by calculating the execution times of four benchmarks using \emph{dataClay} without any prefetching, and with \acs{CAPre}. We also compared \acs{CAPre} with the \emph{Referenced-Objects Predictor (ROP)}, defined in Section \ref{sec:intro}, using different \emph{fetch depths}, which indicate the levels of related objects that the ROP should prefetch. 

For each experiment, we executed the benchmark 10 times and took the average execution times. We ran all of the experiments on a cluster of 5 nodes interconnected by a 10GbE link. Each node is composed of a 4-core Intel Xeon E5-2609v2 processor (2.50GHz), a 32GB DRAM (DDR3) and a 1TB HDD (WD10JPVX 5400rpm). We deployed \emph{dataClay} on the cluster using one node as both the client and Logic Module, and 4 nodes as 4 distinct Data Services.

The rest of this section exposes the results of our experiments on each of the studied benchmarks separately.

\subsubsection{OO7}
\begin{figure}[!b]
\centering
\includegraphics[width=0.6\textwidth]{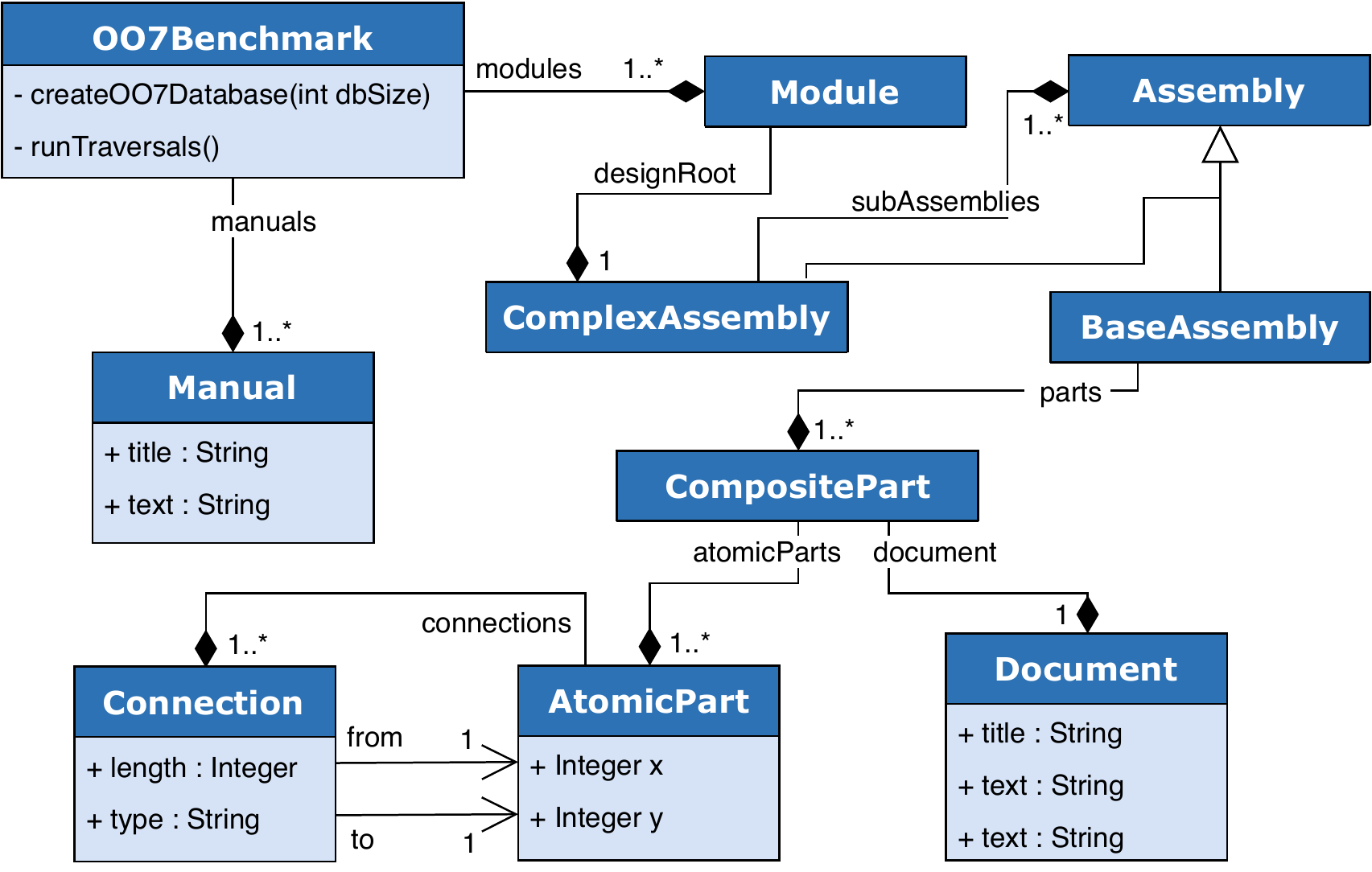}
\caption{Class diagram of the OO7 benchmark.}
\label{fig:oo7_class_diagram}
\end{figure}
OO7 is the \emph{de facto} standard benchmark for POSs and object-oriented databases \cite{oo7Carey}. Its data model is meant to be an abstraction of different CAD/CAM/CASE applications and contains a recursive data structure involving a set of classes with complex inheritance and composition relationships, as depicted in Figure \ref{fig:oo7_class_diagram}. The benchmark includes a random data generator that takes as parameter the size of the database to be generated: small (\textasciitilde1,000 objects), medium (\textasciitilde30,000 objects) and large (\textasciitilde600,000 objects). The benchmark also has an implemented set of 6 traversals, from which we executed the following:
\begin{itemize}
    \item \emph{t1:} tests the data access speed by traversing the benchmark's data model starting from the object \emph{Module}.
    \item \emph{t2a, t2b} and \emph{t2c:} test the update speed by updating different numbers of \emph{Composite Parts} and \emph{Atomic Parts}.
\end{itemize}
We did not execute the two remaining traversals, \emph{t8} and \emph{t9}, given that they were designed to test text processing speed and only load one persistent object, \emph{Manual}.

\begin{figure*}[!t]
\centering
    \begin{subfigure}[b]{\textwidth}
        \centering
        \includegraphics[width=\textwidth,trim=0cm 0cm 0cm 0.2cm ,clip]{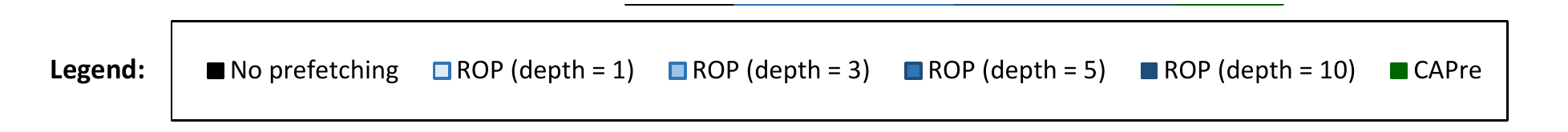}
    \end{subfigure}
    \begin{subfigure}[b]{\textwidth}
        \centering
        \begin{subfigure}[b]{0.3\textwidth}
            \centering
       	\includegraphics[width=\textwidth]{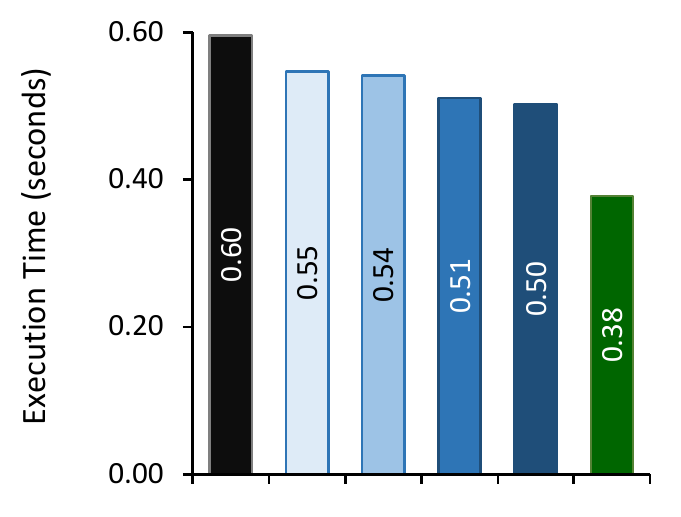}
            \caption*{$ small\ DB $}
            \label{fig:oo7_multiDS_exec_0}
        \end{subfigure}
        \begin{subfigure}[b]{0.3\textwidth}
            \centering
        	\includegraphics[width=\textwidth]{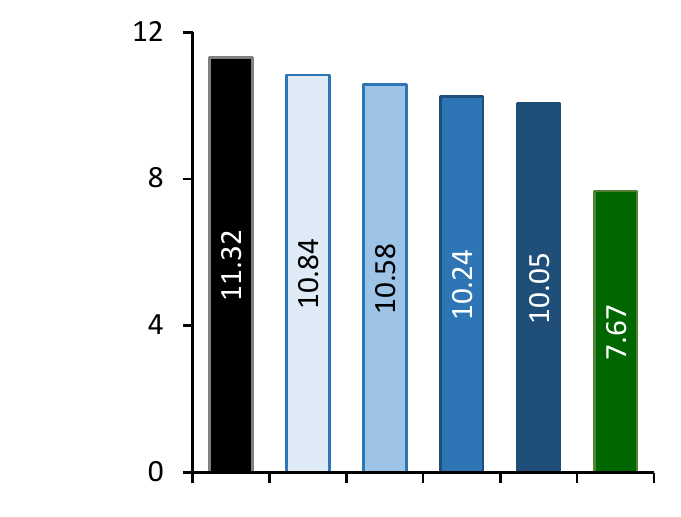}
            \caption*{$ medium\ DB $}
            \label{fig:oo7_multiDS_exec_1}
        \end{subfigure}
        \begin{subfigure}[b]{0.3\textwidth}
            \centering
            \includegraphics[width=\textwidth]{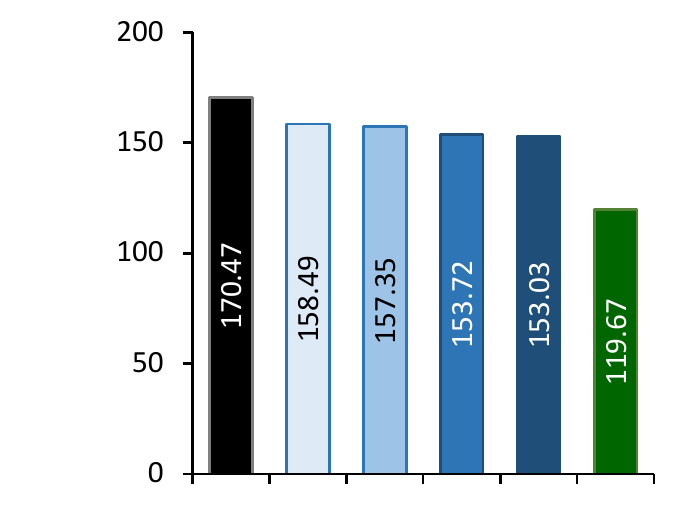}
            \caption*{$ large\ DB $}
            \label{fig:oo7_multiDS_exec_2}
        \end{subfigure}
        \caption{Traversal \emph{t1}}
        \label{fig:oo7_t1_exec_times}
    \end{subfigure}
    \begin{subfigure}[b]{\textwidth}
        \centering
        \begin{subfigure}[b]{0.3\textwidth}
            \centering
        	\includegraphics[width=\textwidth]{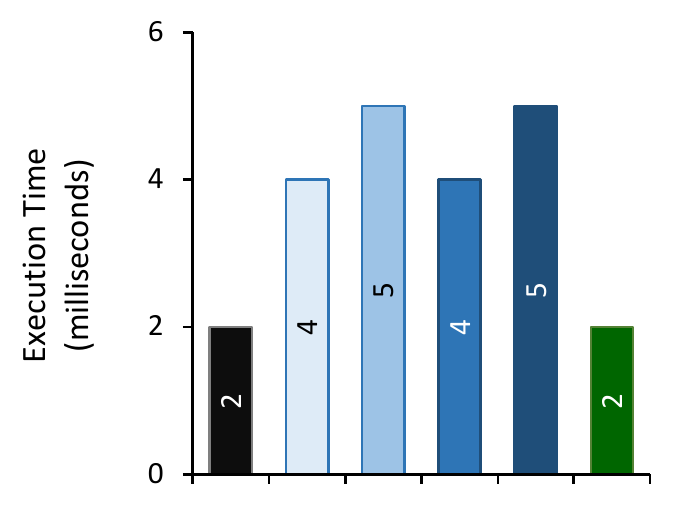}
            \caption*{$ small\ DB $}
            \label{fig:oo7_t2b_exec_0}
        \end{subfigure}
        \begin{subfigure}[b]{0.3\textwidth}
            \centering
        	\includegraphics[width=\textwidth]{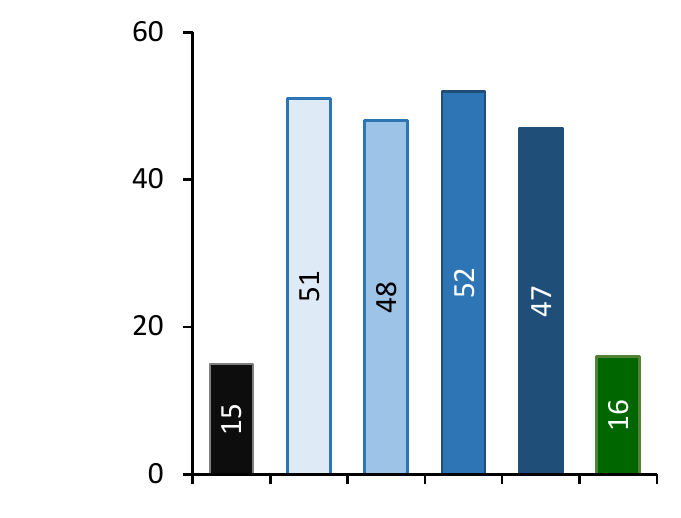}
            \caption*{$ medium\ DB $}
            \label{fig:oo7_t2b_exec_1}
        \end{subfigure}
        \begin{subfigure}[b]{0.3\textwidth}
            \centering
            \includegraphics[width=\textwidth]{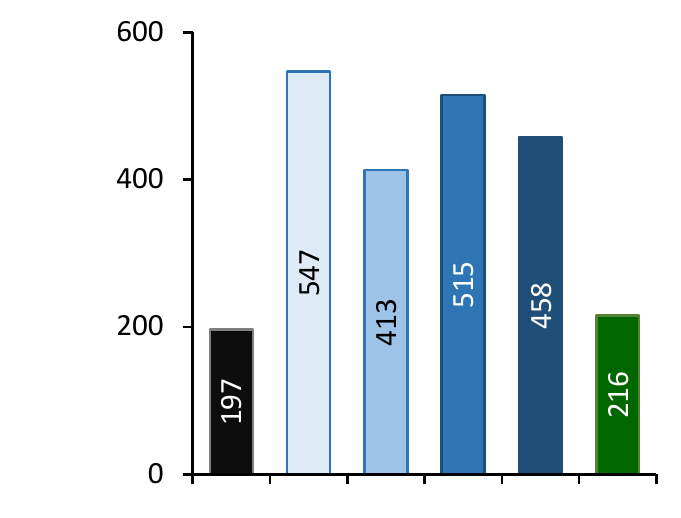}
            \caption*{$ large\ DB $}
            \label{fig:oo7_t2b_exec_2}
        \end{subfigure}
        \caption{Traversal \emph{t2b}}
        \label{fig:oo7_t2b_exec_times}
    \end{subfigure}
\caption{Execution times of the traversals \emph{t1} and \emph{t2b} of the OO7 benchmark.}
\label{fig:oo7_exec_times}
\end{figure*}

Figure \ref{fig:oo7_t1_exec_times} shows the execution times of the traversal \emph{t1} with the three OO7 database sizes. It indicates that \acs{CAPre} offers more improvement to the original execution time than the ROP, which offers gradually better improvement when increasing its fetch depth from 1 to 5 before it stagnates with a fetch depth of 10. This behavior is expected since ROP can only prefetch objects up to a certain depth before running out of referenced objects to prefetch. On the other hand, \acs{CAPre} does not depend on a predefined fetch depth and can prefetch as many levels of related objects as predicted by the code analysis it performs. In addition, given that \acs{CAPre} is able to know which collections will be accessed, their elements can also be prefetched, something that is not done by the ROP algorithm regardless of its depth (prefetching a collection that may not be used is too much overhead). This enables \acs{CAPre} to prefetch many more objects, and thus take more benefit from the parallel access to the distributed storage.

When considering previous work on prefetching that have used OO7 as a benchmark, Ibrahim \emph{et al.} report an improvement of 7\% in execution time with the small OO7 database while Bernstein \emph{et al.} report an improvement of 11\%  on the medium-sized database \cite{bernstein1999context}. While these numbers are not directly comparable to the ones obtained in our experiments given that the approaches use a different POS, with different levels of optimization and run their experiments on different hardware, it is worth mentioning that \acs{CAPre} achieves an improvement of 30\% and 26\% with the small and medium OO7 databases respectively.

As for the traversal \emph{t2b}, Figure \ref{fig:oo7_t2b_exec_times} shows that neither \acs{CAPre} nor the ROP offer any improvement, since the latency of the traversal is not caused by data access but rather by the time taken to store the updated objects. However, the figure also indicates that using the ROP produces significant overhead, caused by the fact that it prefetches the objects referenced from the object being updated, when in fact these objects are never accessed. By contrast, \acs{CAPre} does not prefetch these objects since it takes into consideration the application's code and is aware that they are not needed, thus producing very little overhead. Note that the execution times of the traversals \emph{t2a} and \emph{t2c} were left out of this paper because they exhibit similar behavior in terms of added overhead for both \acs{CAPre} and the ROP.

\subsubsection{Wordcount}
\begin{figure}[!t]
\centering
\includegraphics[width=0.6\textwidth]{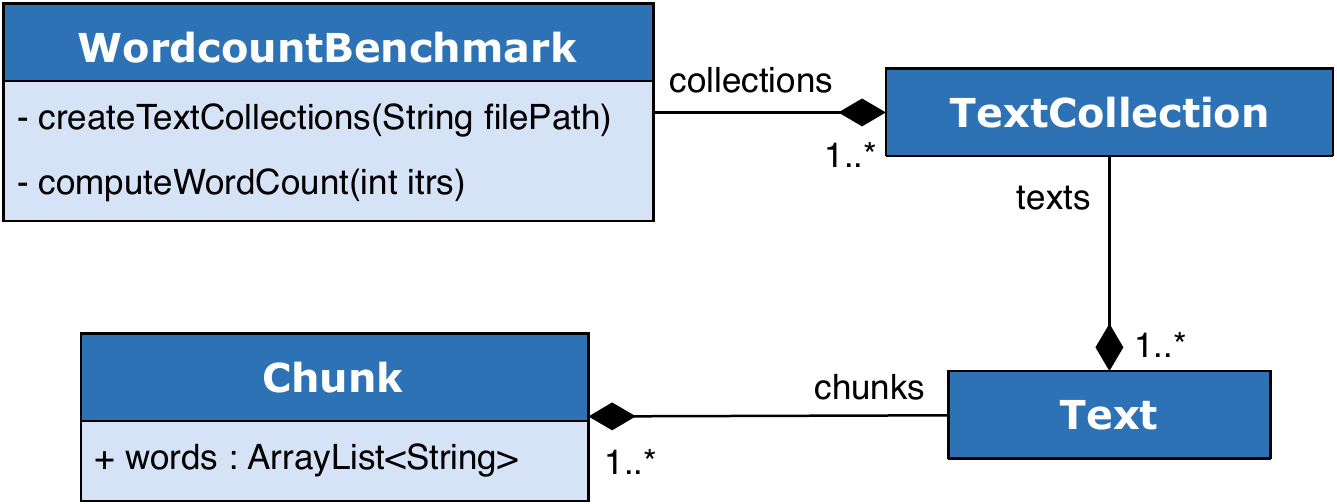}
\caption{Class diagram of the Wordcount benchmark.}
\label{fig:wordcount_class_diagram}
\end{figure}

Wordcount is a parallel algorithm that parses input text files, splitting their text lines into words, and outputs the number of appearances of each unique word. Due to the resemblance of this algorithm to the problem of creating histograms, Wordcount is commonly used as a Big Data benchmark. Unlike OO7, the data model of this benchmark, depicted in Figure \ref{fig:wordcount_class_diagram}, is fairly simple. It consists of several \emph{Text Collections}, each containing one or more \emph{Texts} representing the input files. Each of the \emph{Text} objects in turn contains one or more \emph{Chunks}, which represent fragments of the text, and contain the words to be counted.

In our experiments, we used a data set of 8 files, containing a total of $ 10^7 $ words, divided them into four collections, and distributed the collections among the four \emph{dataClay} Data Services. Furthermore, we ran the benchmark with different numbers of chunks $ c $, ranging from one chunk containing all the words in each text (i.e. few large objects) to $ 10^6 $ chunks per text containing very few words (i.e. many small objects).

\begin{figure*}[!t]
\centering
    \begin{subfigure}[b]{\textwidth}
        \centering
        \includegraphics[width=\textwidth,trim=0cm 0cm 0cm 0.2cm ,clip]{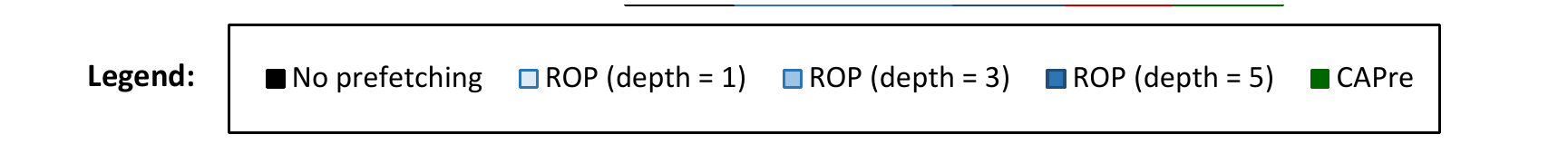}
    \end{subfigure}
    \begin{subfigure}[b]{0.24\textwidth}
        \centering
    	\includegraphics[width=\textwidth]{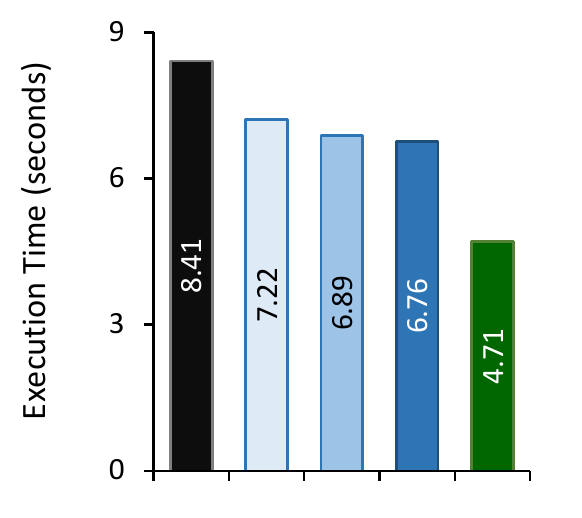}
        \caption*{$ c = 1 $}
	    \label{fig:wordcount_multiDS_exec_0}
    \end{subfigure}
    \begin{subfigure}[b]{0.24\textwidth}
        \centering
    	\includegraphics[width=\textwidth]{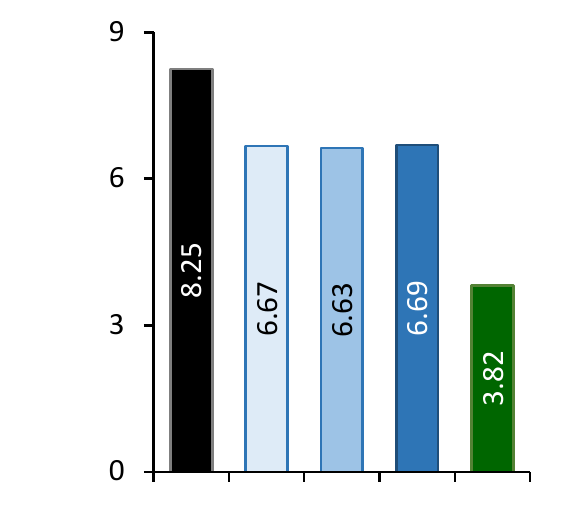}
        \caption*{$ c = 10 $}
	    \label{fig:wordcount_multiDS_exec_1}
    \end{subfigure}
    \begin{subfigure}[b]{0.24\textwidth}
        \centering
	    \includegraphics[width=\textwidth]{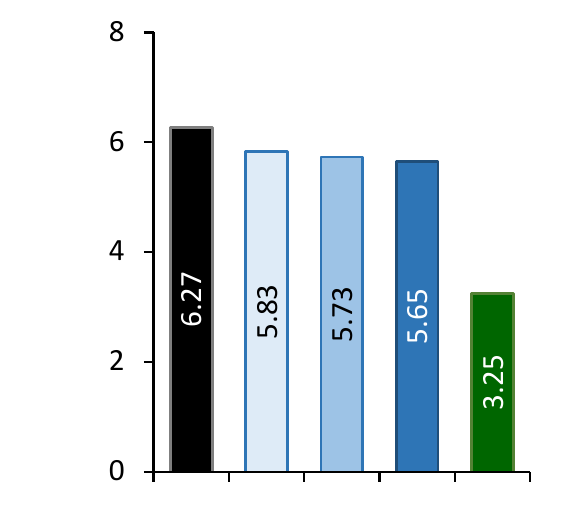}
        \caption*{$ c = 10^2 $}
	    \label{fig:wordcount_multiDS_exec_2}
    \end{subfigure}
    \begin{subfigure}[b]{0.24\textwidth}
        \centering
    	\includegraphics[width=\textwidth]{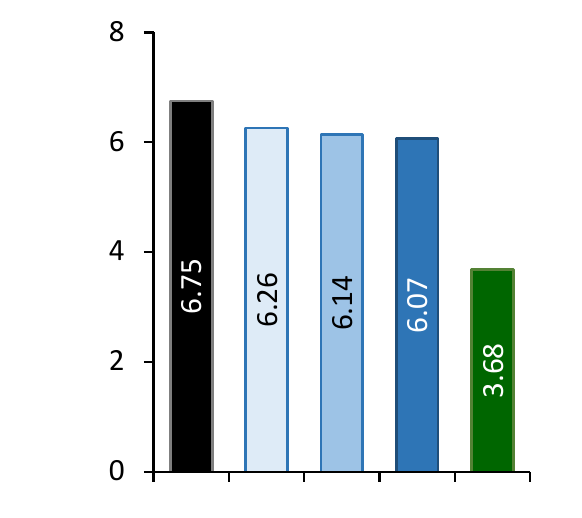}
        \caption*{$ c = 10^3 $}
	    \label{fig:wordcount_multiDS_exec_3}
    \end{subfigure}
    \begin{subfigure}[b]{0.24\textwidth}
        \centering
    	\includegraphics[width=\textwidth]{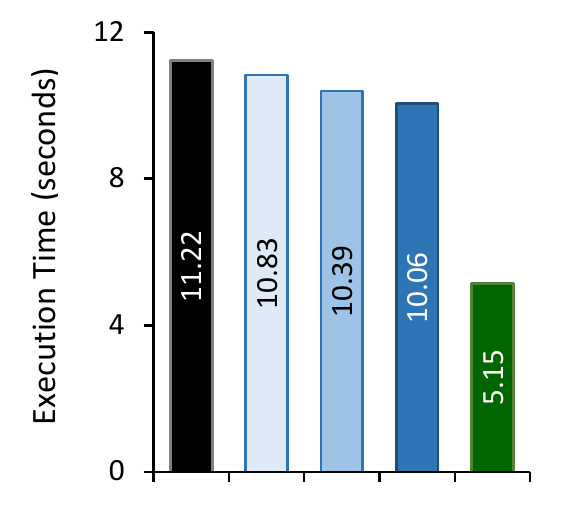}
        \caption*{$ c = 10^4 $}
	    \label{fig:wordcount_multiDS_exec_4}
    \end{subfigure}
    \begin{subfigure}[b]{0.24\textwidth}
        \centering
	    \includegraphics[width=\textwidth]{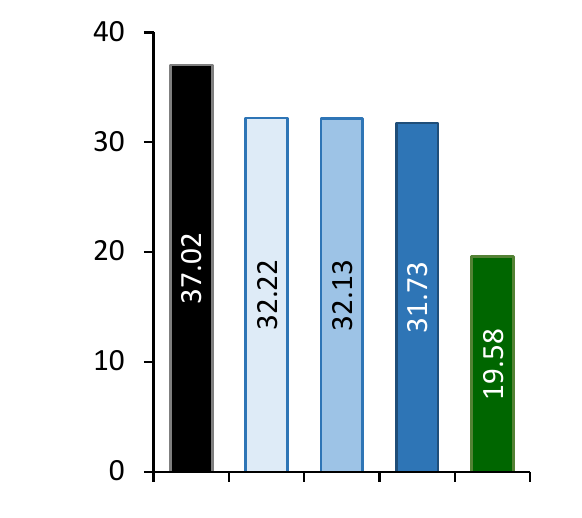}
        \caption*{$ c = 10^5 $}
	    \label{fig:wordcount_multiDS_exec_5}
    \end{subfigure}
    \begin{subfigure}[b]{0.24\textwidth}
        \centering
    	\includegraphics[width=\textwidth]{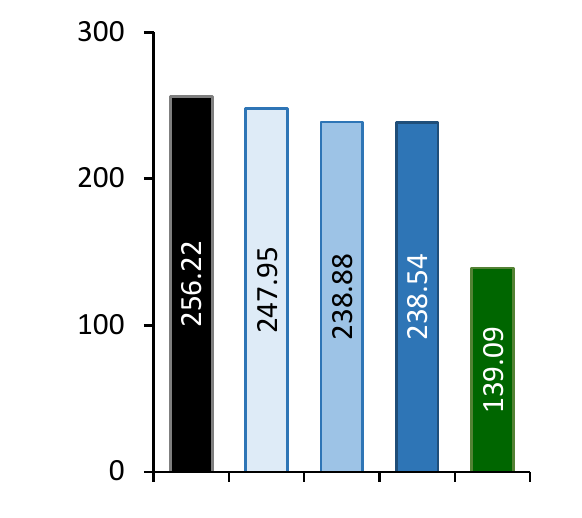}
        \caption*{$ c = 10^6 $}
	    \label{fig:wordcount_multiDS_exec_6}
    \end{subfigure}
\caption{Execution times of the Wordcount benchmark.}
\label{fig:wordcount_exec_times}
\end{figure*}

Figure \ref{fig:wordcount_exec_times} shows the execution times of the Wordcount benchmark. Given that the data model of Wordcount is simpler than OO7, we can see that the ROP stagnates at a lower fetch depth of 3.
For this motivation, we do not include the results for ROP with a fetch depth of 10 with any of the rest of experiments in this section. On the other hand, given that most of the data are collections, \acs{CAPre} knows which ones to prefetch and thus does brings them to main memory (something that, as we have mentioned cannot be done by ROP) increasing the hit ratio and, thus, reducing the execution time by more than 50\% in some cases.

This improvement is considerably higher than what we obtained with OO7, because the Wordcount data model contains many collection associations, which can be prefetched by our approach. Finally, Figure \ref{fig:wordcount_exec_times} also shows that \acs{CAPre} offers stable improvement regardless of the number of chunks, which indicates that it can be equally beneficial for applications that handle a small number of large objects or many small-sized objects.

\subsubsection{K-Means}

\begin{figure}[!t]
\centering
\includegraphics[width=0.6\textwidth]{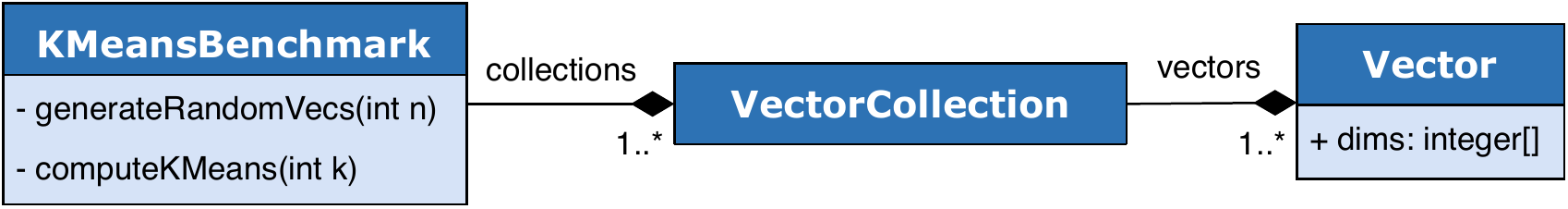}
\caption{Class diagram of the K-Means benchmark.}
\label{fig:kmeans_class_diagram}
\end{figure}

K-Means is a clustering algorithm commonly used as a Big Data benchmark that aims to partition $ n $ input vectors into $ k $ clusters in which each vector belongs to the cluster with the nearest mean. It is a complex recursive algorithm that requires several iterations to reach a converging solution. The data model of K-Means that we used, depicted in Figure \ref{fig:kmeans_class_diagram}, consists of a set of \emph{VectorCollections} each containing a subset of the $ n $ input \emph{Vectors}. We ran our experiments using various numbers of randomly generated vectors, $ n $, each consisting of 10 dimensions, and different values of $ k $. We also divided the input vectors into 4 collections and distributed the collections among the \emph{dataClay} Data Services.

\begin{figure*}[!t]
\centering
    \begin{subfigure}[b]{\textwidth}
        \centering
        \includegraphics[width=\textwidth,trim=0cm 0cm 0cm 0.2cm ,clip]{figures/wordcount_exec_times_legend.pdf}
    \end{subfigure}
    \begin{subfigure}[b]{0.24\textwidth}
        \centering
    	\includegraphics[width=\textwidth]{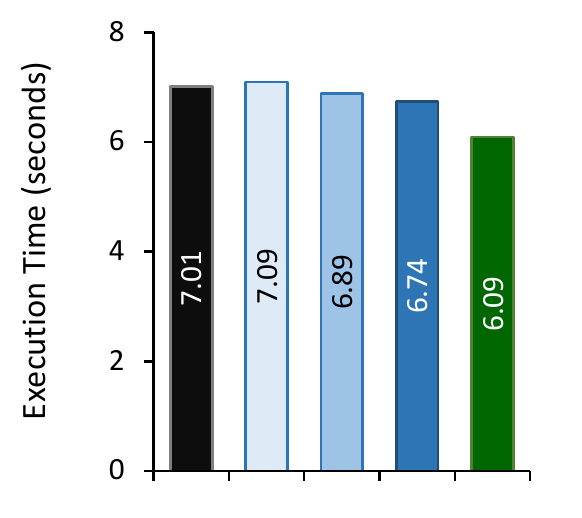}
        \caption*{$ n = 10^3,\ k = 4 $}
	    \label{fig:kmeans_exec_0}
    \end{subfigure}
    \begin{subfigure}[b]{0.24\textwidth}
        \centering
    	\includegraphics[width=\textwidth]{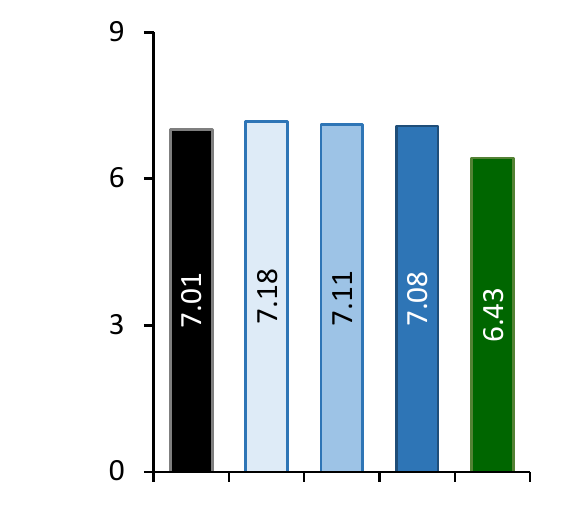}
        \caption*{$ n = 10^4,\ k = 4 $}
	    \label{fig:kmeans_exec_1}
    \end{subfigure}
    \begin{subfigure}[b]{0.24\textwidth}
        \centering
    	\includegraphics[width=\textwidth]{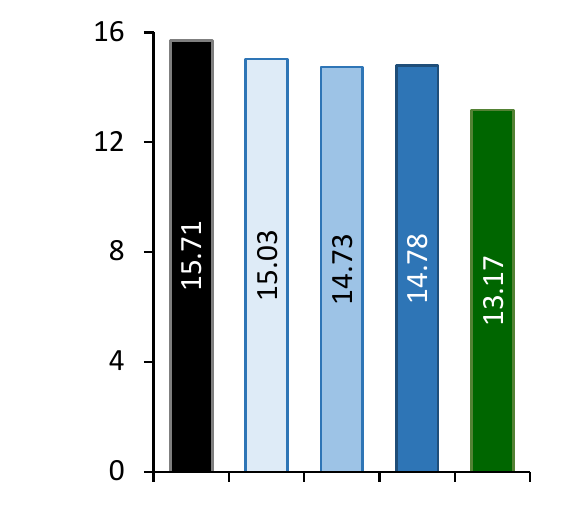}
        \caption*{$ n = 10^5,\ k = 40 $}
	    \label{fig:kmeans_exec_2}
    \end{subfigure}
    \begin{subfigure}[b]{0.24\textwidth}
        \centering
    	\includegraphics[width=\textwidth]{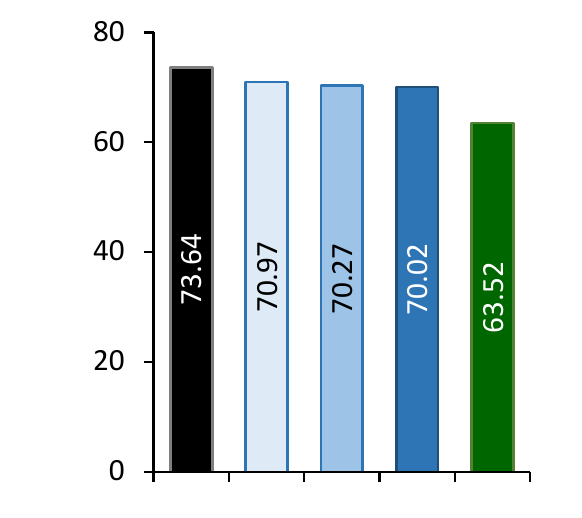}
        \caption*{$ n = 10^6,\ k = 400 $}
	    \label{fig:kmeans_exec_3}
    \end{subfigure}
\caption{Execution times of the K-Means benchmark.}
\label{fig:kmeans_exec_times}
\end{figure*}

Figure \ref{fig:kmeans_exec_times} shows the execution times of this benchmark. In this case, the ROP does not offer any significant improvement regardless of the fetch depth given that the benchmark's data model does not contain any single associations that can be prefetched. On the contrary, \acs{CAPre}  achieves better improvement, reducing between 9\% and 15\% of the benchmark's execution time, when prefetching data collections in parallel, which again shows the advantage of \acs{CAPre}.

\subsubsection{Princeton Graph Algorithms}

\begin{figure}[t]
\centering
\includegraphics[width=0.6\textwidth]{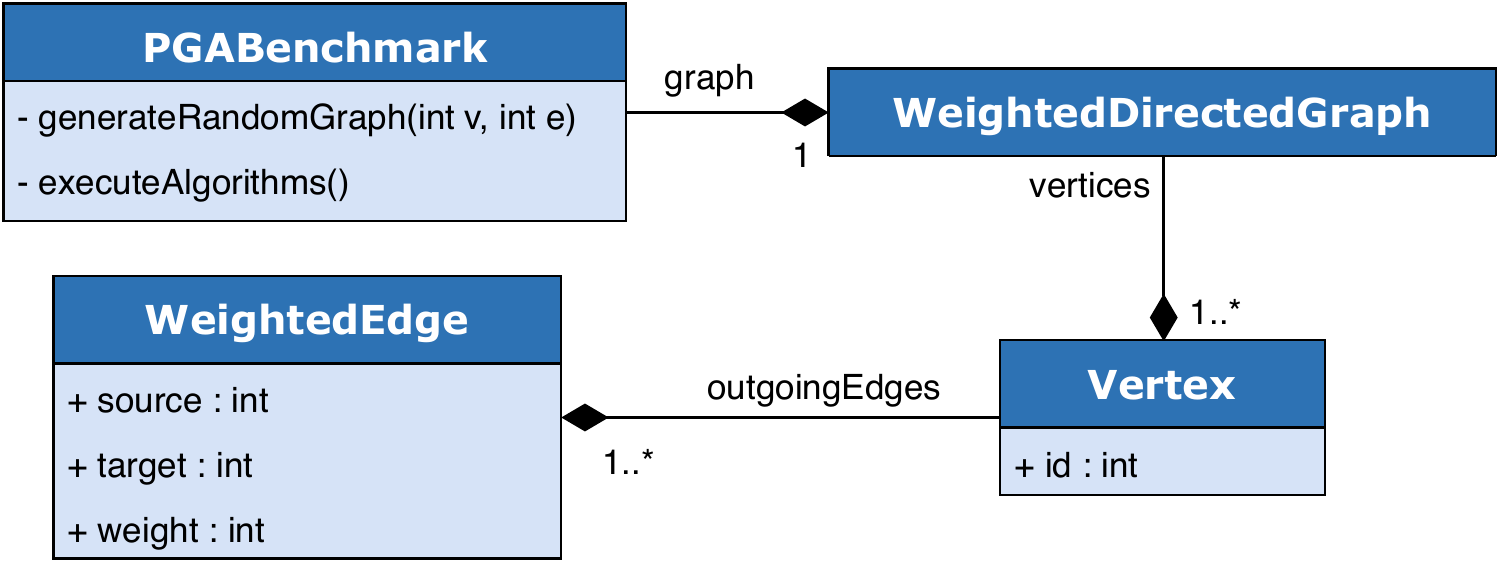}
\caption{Class diagram of the PGA benchmark.}
\label{fig:pga_class_diagram}
\end{figure}

\begin{figure*}[t]
\centering
    \begin{subfigure}[b]{\textwidth}
        \centering
        \includegraphics[width=\textwidth,trim=0cm 0cm 0cm 0.2cm ,clip]{figures/wordcount_exec_times_legend.pdf}
    \end{subfigure}
    \begin{subfigure}[b]{\textwidth}
        \centering
        \begin{subfigure}[b]{0.24\textwidth}
            \centering
        	\includegraphics[width=\textwidth]{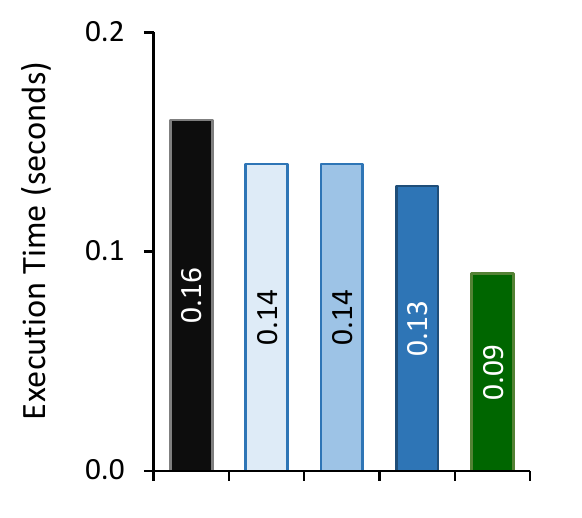}
            \caption*{$ v = 10^2,\ e = 10^3 $}
            \label{fig:pga_dfs_exec_0}
        \end{subfigure}
        \begin{subfigure}[b]{0.24\textwidth}
            \centering
        	\includegraphics[width=\textwidth]{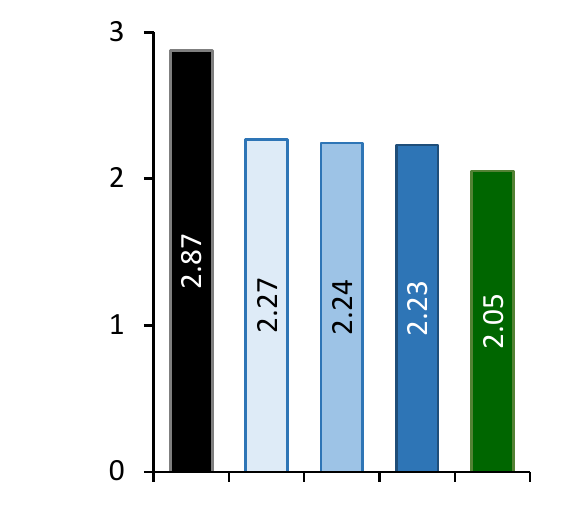}
            \caption*{$ v = 10^3,\ e = 10^4 $}
            \label{fig:pga_dfs_exec_1}
        \end{subfigure}
        \begin{subfigure}[b]{0.24\textwidth}
            \centering
        	\includegraphics[width=\textwidth]{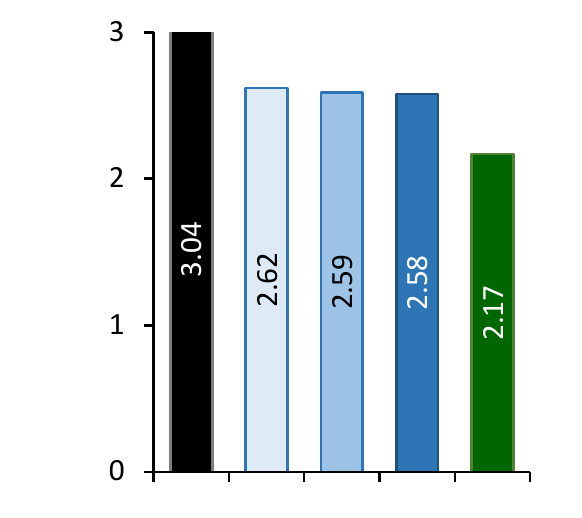}
            \caption*{$ v = 10^3,\ e = 10^5 $}
            \label{fig:pga_dfs_exec_2}
        \end{subfigure}
        \begin{subfigure}[b]{0.24\textwidth}
            \centering
        	\includegraphics[width=\textwidth]{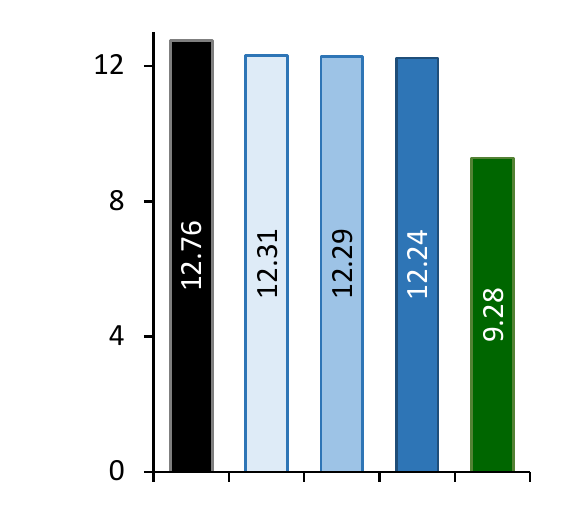}
            \caption*{$ v = 10^4,\ e = 10^5 $}
            \label{fig:pga_dfs_exec_3}
        \end{subfigure}
        \caption{Depth-First Search (DFS)}
        \label{fig:pga_dfs_exec_times}
    \end{subfigure}
    \begin{subfigure}[b]{\textwidth}
        \centering
        \begin{subfigure}[b]{0.24\textwidth}
            \centering
        	\includegraphics[width=\textwidth]{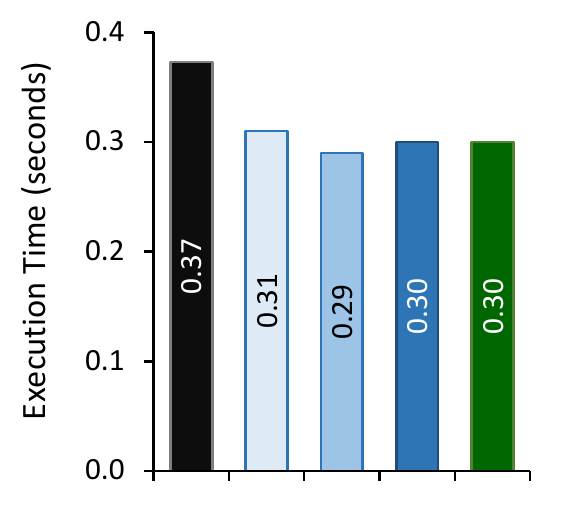}
            \caption*{$ v = 10^2,\ e = 10^3 $}
            \label{fig:pga_bf_exec_0}
        \end{subfigure}
        \begin{subfigure}[b]{0.24\textwidth}
            \centering
        	\includegraphics[width=\textwidth]{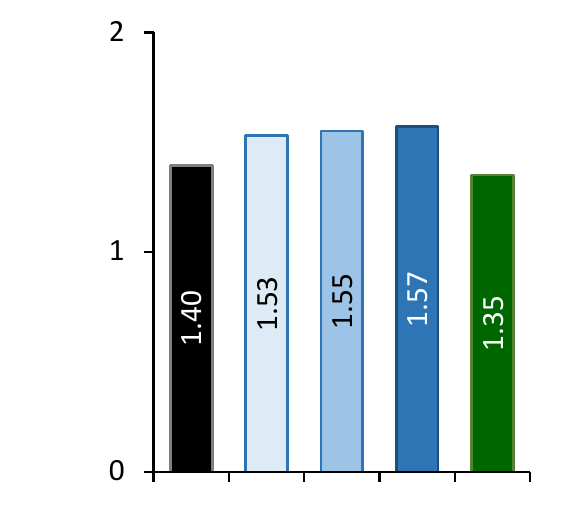}
            \caption*{$ v = 10^3,\ e = 10^4 $}
            \label{fig:pga_bf_exec_1}
        \end{subfigure}
        \begin{subfigure}[b]{0.24\textwidth}
            \centering
        	\includegraphics[width=\textwidth]{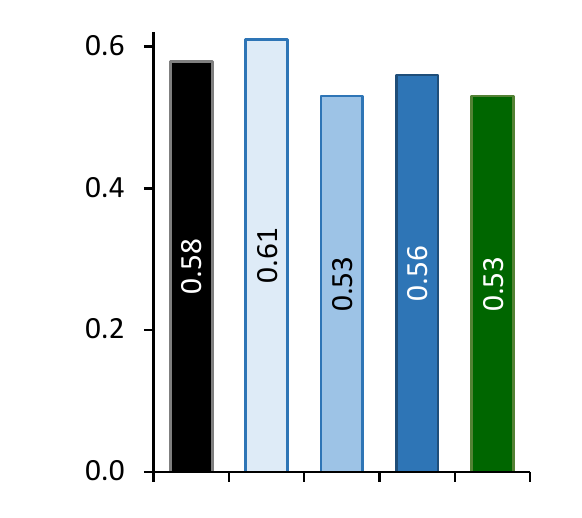}
            \caption*{$ v = 10^3,\ e = 10^5 $}
            \label{fig:pga_bf_exec_2}
        \end{subfigure}
        \begin{subfigure}[b]{0.24\textwidth}
            \centering
        	\includegraphics[width=\textwidth]{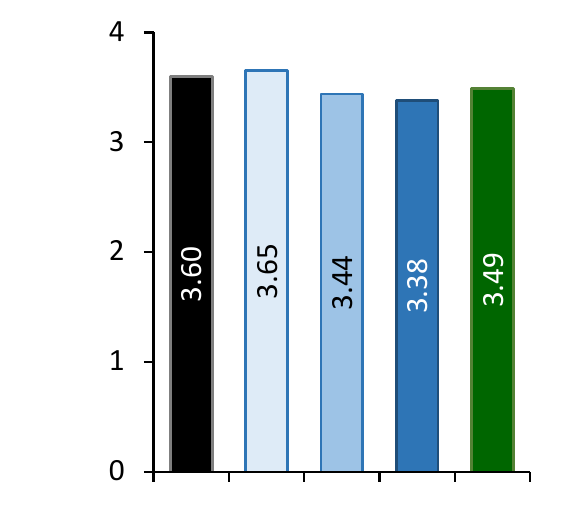}
            \caption*{$ v = 10^4,\ e = 10^5 $}
            \label{fig:pga_bf_exec_3}
        \end{subfigure}
        \caption{Bellman-Ford Shortest Path}
        \label{fig:pga_bf_exec_times}
    \end{subfigure}
\caption{Execution times of the Princeton Graph Algorithms benchmark.}
\label{fig:pga_exec_times}
\end{figure*}


The Princeton Graph Algorithms (PGA) is a benchmark used to test the execution times of complex graph traversal algorithms using different types of graphs (e.g. undirected, directed, weighted) \cite{pgaBenchmark}. Figure \ref{fig:pga_class_diagram} depicts the subset of the benchmark's classes that we used in our experiments. Namely, we executed the Depth-First Search (DFS) and Bellman-Ford Shortest Path algorithms using a \emph{WeightedDirectedGraph}. The graph consists of a set of \emph{Vertex} objects, each containing the outgoing \emph{WeightedEdges} of the vertex. We ran our experiments using different numbers of randomly generated vertices $ v $ and edges $ e $, which we chose to construct graphs with different levels of edge density. As with the rest of the benchmarks, we distributed the data among the four Data Services of \emph{dataClay}.

Figure \ref{fig:pga_dfs_exec_times} shows that the execution times of the DFS algorithm are similar to those reported for the WordCount benchmark; where \acs{CAPre} doubles the improvement achieved by ROP and the same rationale applies. On the other hand, Figure \ref{fig:pga_bf_exec_times} indicates that even when using \acs{CAPre}, we do not see significant improvement in the execution time of the Bellman-Ford algorithm. This is due to the fact that this algorithm does not access the graph's vertices in a predetermined order, but rather starts from a source vertex and applies a trial-and-error approach to reach the shortest path solution using various intermediate data structures, and thus predicting access to the objects it uses is more difficult. Nevertheless, it is also important to notice that in these cases, \acs{CAPre} knows what not to prefetch and does not add unnecessary overhead as it happens in some cases with ROP.

\subsection{Discussion}

The results obtained from our experiments indicate that \acs{CAPre} offers the highest improvement in execution time when used with applications with a complex data model, such as OO7. This is due to the fact that CAPre is based on type graphs, which analyze the way that the data model of the application is accessed by its methods. As such, the more complex a data model is the more information on which to base the prefetching predictions can be retrieved.

Nonetheless, the fact that \acs{CAPre} can safely predict access to collections as well as single objects, allows it to be used with simple data models that contain many collection associations as well, such as the case with the Wordcount and K-Means benchmarks. This prediction of access to collection also increases the amount of objects to be prefetched at a time, thus giving \acs{CAPre} more margin to take advantage of any potential parallelism in the POS when prefetching the predicted objects.

This prediction of access to collections of persistent objects, and the associated parallel prefetching of these objects, is an important area where \acs{CAPre} outperforms ROP. As discussed throughout this section, ROP is limited to predicting access to single objects and unable to predict access to collections, due to its heuristic of retrieving objects related to the one currently being accessed. This in turn means that a prefetching system based on ROP is not able to take advantage of parallelism in the POS, given that collections of objects that can be accessed in parallel are never predicted for prefetching.

In terms of data size, the experiments indicate that \acs{CAPre} provides the same level of improvement regardless of the number or size of persistent objects manipulated by each benchmark. This indicates that CAPre can be used with both applications that manipulate a large number of small persistent objects, as well as with those that manipulate a small number of large persistent objects. When compared with the ROP, \acs{CAPre} achieves at least the same improvement and, in cases where prefetching is not needed, the negative effect on application performance is significantly smaller than when using ROP.

Throughout our experiments, we encountered one limitation of \acs{CAPre}, with the Bellman-Ford shortest path algorithm, where it could not offer significant improvement because the algorithm accesses persistent objects in a random order that is difficult to predict. Theoretically, we can also run into another limitation when the objects accessed by different branches of a conditional statement do not have any overlap. In this case, \acs{CAPre} would retrieve many unnecessary objects given that it prefetches the objects predicted by the union of the prefetching hints of the different branches. However, our analysis of the SF110 corpus, detailed in Section \ref{sec:bdnom}, shows that this limitation only occurs in a very small minority of the analyzed applications, and that in the majority of cases there is a big overlap between the objects accessed by different branches of a conditional statement (even though the methods executed on these objects may be very different).
 

In these cases, any prefetching approach that uses a compile-time prediction technique will face the problem of unpredictability of the accessed objects, as evident by the inability of ROP to offer any improvement in the execution time as well. One solution to this problem is to use a hybrid approach that collects some information during runtime in order to complement the predictions made prior to the execution of the application. Such an approach will evidently have to be studied and analyzed in detail in order to determine the overhead that it might introduce.

Finally, \acs{CAPre} currently uses the Java Virtual Machine’s (JVM) predefined threadpool to execute the parallel prefetching of collections. This approach reduces the costs of creating and destroying threads and delegates the management of the threads to the JVM. Nonetheless, it does not allow us to test the effects that the number of threads has on the experiment results, given that it is the JVM that decides the optimal number of threads to create without overloading the machine. It may be interesting, as future work, to take control of the thread management operations from the JVM in order to evaluate how the number of prefetching threads influences the efficiency of the prefetching performed by \acs{CAPre}.
\section{Conclusions}
\label{sec:conclusions}
In this paper, we presented \acs{CAPre}, a prefetching system for Persistent Object Stores based on static code analysis of object-oriented applications. We detailed the analysis we perform to obtain prefetching hints that predict which persistent objects are accessed by the application and how we use code generation and injection to prefetch the predicted objects when the application is executed. We also optimized the system by parallelizing the generated prefetching methods, allowing objects to be prefetched from various nodes of a distributed POS in parallel. Afterwards, we integrated \acs{CAPre} into a distributed POS and performed a series of experiments on known benchmarks to evaluate the improvement to application performance that it can achieve.

In the future, we want to address cases where \acs{CAPre} offers limited improvement by collecting more information during application execution, while studying the overhead that such a hybrid approach might introduce. We also plan to use the predictions made by the developed static code analysis to apply other performance improvement techniques in conjunction with prefetching, such as smart cache replacement policies \cite{Jaleel2012,Jeong2003,Keramidas2007} and dynamic data placement \cite{Lee2014,Maheshwari2012}.

\section*{Acknowledgements}
This work has been supported by the European Union's Horizon 2020 research and innovation program under the BigStorage European Training Network (ETN) (grant H2020-MSCA-ITN-2014-642963), the Spanish Ministry of Science and Innovation (contract TIN2015-65316) and the Generalitat de Catalunya (contract 2014-SGR-1051).

\section*{Bibliography}

\bibliographystyle{elsarticle-num}
\bibliography{bib/references}

\end{document}